\documentclass[%
 aps,pre,
 amsmath,amssymb,
 reprint,%
]{revtex4-2}

\usepackage{mpsi}
\usepackage{graphicx}
\usepackage{dcolumn}
\usepackage{bm}
\usepackage{amsfonts,amsthm,mathbbol}
\usepackage[utf8]{inputenc}
\usepackage[T1]{fontenc}
\usepackage{mathptmx}
\usepackage{color}
\usepackage{amsmath}
\usepackage{setspace}
\usepackage{ulem}

\usepackage{ragged2e}
\usepackage{pgfplots}
\newcommand{\sgn}{\operatorname{sgn}}

\begin{document}

\title[Short-time expansion of one-dimensional Fokker-Planck equations with heterogeneous diffusion]{Short-time expansion of one-dimensional Fokker-Planck equations with heterogeneous diffusion}

\author{Tom Dupont}%
\email{tom.dupont@centrale.centralelille.fr}  
\affiliation{University of Lille, CNRS, Centrale Lille, Univ. Polytechnique Hauts-de-France, UMR 8520 - IEMN - Institut d'{\'E}lectronique, de Micro{\'e}lectronique et de Nanotechnologie, F-59000 Lille, France
}

\author{Stefano Giordano}%
\email{stefano.giordano@univ-lille.fr}  
\affiliation{University of Lille, CNRS, Centrale Lille, Univ. Polytechnique Hauts-de-France, UMR 8520 - IEMN - Institut d'{\'E}lectronique, de Micro{\'e}lectronique et de Nanotechnologie, F-59000 Lille, France
}

\author{Fabrizio Cleri}%
\email{fabrizio.cleri@univ-lille.fr}  
\affiliation{University of Lille, Institut d'{\'E}lectronique, de Micro{\'e}lectronique et de Nanotechnologie (IEMN CNRS UMR8520) and Departement de Physique, F-59652 Villeneuve d'Ascq, France
}

 \author{Ralf Blossey}
\email{ralf.blossey@univ-lille.fr}
\affiliation{University of Lille, Unit{\'e} de Glycobiologie Structurale et Fonctionnelle (UGSF), CNRS UMR8576, F-59000 Lille, France}

\date{\today}

\begin{abstract}
We formulate a short-time expansion for one-dimensional Fokker-Planck equations with spatially dependent diffusion coefficients, derived from stochastic processes with Gaussian 
white noise, for general values of the discretization parameter $0 \leq \alpha \leq 1$ of the stochastic integral. The kernel of the Fokker-Planck equation (the propagator) can be expressed as a product of a singular and a regular term. While the singular term can be given in closed form, the regular term can be computed from a Taylor expansion whose coefficients obey simple ordinary differential equations. We illustrate the application of our approach with examples taken from statistical physics and biophysics. Further, we show how our formalism allows to define a class of stochastic equations which can be treated exactly. The convergence of the expansion cannot be guaranteed independently from the discretization parameter $\alpha$.
\end{abstract}

\maketitle


\section{Introduction}

The Fokker-Planck (FP) equation is one of the most important computational tools in statistical mechanics to obtain an approximation of the probability distribution in time and space of a generic system including a stochastic driving force. As such, the FP equation has found numerous applications in widely disparate domains, ranging from plasma physics, biophysics, physical chemistry, finance, and others \cite{risken1989,gardiner2009,coffey2004}. 
Interestingly, the underlying theory of this
linear parabolic partial differential equation is still not yet fully developed, and continues to hold surprises of physical relevance. Recent developments e.g. concern spatially varying diffusivity adopted to describe anomalous diffusion \cite{Silva2011,Cherstvy2013,Metzler2014,Cherstvy2017,Wang2020,
Ritschel2021,Vinod2022,Vinod2022b,ribeiro2023}. Among the many statistical properties of heterogeneous diffusion processes studied are in particular non-stationarity and non-ergodicity; see e.g. results on the non-existence of normalizable stationary solutions, or equilibrium probability distributions, which can be tackled 
by the methods of infinite ergodicity theory \cite{leibovich2019,aghion2019,aghion2020,giordano2023}. 
In particular these latter results show a sensitive dependence on the discretization parameter $\alpha$ of the stochastic integrals in the Langevin equation underlying the FP equation. The two most commonly chosen discretizations are the Itô and Fisk-Stratonovich, using
values of $\alpha$=0 or 1/2, respectively.

These recent results raise the question whether a similar dependence on the discretization parameter $\alpha$ exists when considering the short-time behaviour of the FP equation, as it can be deduced from systematic expansions in this limit. 
Short-time expansions have been considered before
in the literature \cite{donoso1999,drozdov1993a,drozdov1993,drozdov1995,weiss1995,drozdov1996,nistor2010,nistor2011,bilal2020}, however, to the best of our knowledge, not with this specific question in mind. 

Apart from this important formal aspect, our work was also originally motivated by a specific biophysical problem, namely statistical physics models for molecular motors, in which FP equations with heterogeneous diffusion terms arise \cite{kumar2008,baule2008}. In this context we are particularly interested in chromatin remodelers, molecular motors and enzymes that move and remove nucleosomes in eukaryotic DNA \cite{blossey2019}. We will treat this problem as one of our applications of our approach in this paper.

It is interesting to point out that multiplicative noise has played an essential role in modeling a number of physical and biological phenomena among which we can mention the transmission of signals in  neuron models \cite{bauermann2019,zhu2021}, the phenotypic variability
and gene expression \cite{liu2004,frigola2012}, the stochastic thermodynamics of holonomic systems \cite{manca2016,giordano2019,giordano2021}, the ballistic-to-diffusive transition of the heat propagation \cite{landi2014,palla2020}, and the statistical theory of turbulence \cite{fuchs2022,birnir2013}. Given the widespread use of models based on multiplicative noise and heterogeneous diffusion, it is important to have exact or approximate techniques to solve the corresponding Langevin and Fokker-Planck equations and thus trace the statistical properties of the stochastic processes involved. The approach proposed in this paper, concerning a development of the propagator for short timescales, thus fits into this context and is complementary to all other solution techniques proposed in the literature.

Our paper is organized as follows. We group our material into 
three sections. Section \ref{sechete} begins with a precise
formulation of the mathematical problem we discuss and the
subsequent derivation of the equations of our formal method.
Section \ref{secapp} presents the explicit discussion of four exemplary
stochastic processes, which illustrates the application of our
method with concrete examples. In Section \ref{secfin} we turn our discussion
around by exploiting the properties of our expansion in order to 
find stochastic processes that are explicitly solvable.
We conclude in Section \ref{secconc}  with an outlook on further possible developments.

\section{The short-range expansion of the Fokker-Planck equation 
with heterogeneous diffusion}
\label{sechete}

In this Section, will develop the formalism to study the short-time behavior of the one-dimensional Fokker-Planck equation. We initially state the problem under investigation and then we subdivide the development into two steps: (i) firstly, we obtain the singular short-time part of the propagator, and (ii) secondly we get the differential equations for the regular larger-time part. Finally, we obtain the formal  solution of the hierarchy of differential equations describing the Taylor coefficients of the regular part of the propagator, particularly useful for the applications.

\subsection{Mathematical formulation of the problem}

The starting point of our study is the Langevin equation,
a stochastic differential equation of the form
\begin{equation} 
    \frac{\d x}{\d t}=h(x,t)+g(x,t)\xi(t),
    \label{lan}
\end{equation}
where $x(t)$ is the stochastic process under investigation, and $h$ and $g$ are regular functions. Without limiting generality, we can always assume that $g(x,t)\ge 0$ for any $x\in\mathbb{R}$ and $t\ge t_0$.

Moreover, $\xi(t)$ in Eq. (\ref{lan}) is a Gaussian white noise with average value $\mathbb{E}(\xi(t))=0$, and correlation $\mathbb{E}(\xi(t)\xi(\tau))=2\delta(t-\tau)$, where $\delta(t)$ is the  Dirac delta function. 
Throughout this work, we always consider processes defined on the entire real axis, without any reflecting or absorbing boundary conditions.

For the process $x(t)$, we can introduce the probability density $W(x,t)$. It means that for any real numbers $x$ and $y$ (such that $y\leq x$) we determine the probability that $y\leq x(t)\leq x$ as $\disp\mathbb{P}(y\leq x(t)\leq x)=\int_y^x W(z,t)\d z$, or equivalently, $W(x,t)=\disp\frac{\partial \mathbb{P}(x(t)\leq x)}{\partial x}$. 

In order to completely define the meaning of Eq. (\ref{lan}), we must introduce a parameter $\alpha \in [0,1]$, identifying the integral stochastic interpretation adopted \cite{risken1989,gardiner2009,coffey1985,oksendal2003,coffey2004}. 
Indeed, the value of $\alpha$ defines the position of the point at which we calculate any integrated function in the small intervals of the adopted Riemann sum. This integration rule includes $\alpha=0$ (Itô integral) \cite{ito1950}, $\alpha=\frac{1}{2}$ (Fisk-Stratonovich integral) \cite{fisk1963,stratonovich1966}, and $\alpha=1$ (Hänggi-Klimontovich integral) \cite{haenggi1982,klimontovich1995} (see also Ref. \cite{sokolov2010} for some comparison of these different choices).

The probability density $W(x,t)$ is the solution of the  Fokker-Planck equation \cite{risken1989,gardiner2009,coffey2004,denisov2009,denisov2003,denisov2014}
\begin{equation}
    \frac{\partial W}{\partial t}=-\frac{\partial}{\partial x}\left[\left(h+2\alpha g\frac{\partial g}{\partial x}\right)W\right]+\frac{\partial^2}{\partial x^2}\left(g^2W\right),
\end{equation}
which can be also written in the equivalent form
\begin{equation}
    \frac{\partial W}{\partial t}=\frac{\partial}{\partial x}\left[-hW+g^{2\alpha}\frac{\partial}{\partial x}\left(g^{2(1-\alpha)}W\right)\right].
\end{equation}
The right hand side of this Fokker-Planck equation can be expressed more explicitly as linear combination of  $W$, $\disp\frac{\partial W}{\partial x}$, and $\disp\frac{\partial^2W}{\partial x^2}$. 
We obtain
\begin{equation}
    \frac{\partial W}{\partial t}=-fW-\ell\frac{\partial W}{\partial x}+g^2\frac{\partial^2W}{\partial x^2}
    \label{fp}
\end{equation}
where we introduced the regular functions
\begin{eqnarray}
\label{effe}
    f &=& \frac{\partial h}{\partial x}+2(\alpha-1)\left(\frac{\partial g}{\partial x}\right)^2+2(\alpha-1)g\frac{\partial^2g}{\partial x^2},\\ 
    \label{elle}
    \ell &=& h+2(\alpha-2)g\frac{\partial g}{\partial x}.
\end{eqnarray}
These expressions show that the Fokker-Planck equation, and thus the density $W$, are not affected by the sign of the function $g$, which we assumed as always positive.

To solve the Fokker-Planck equation, we have to specify the initial condition $W(x,t_0)=W_0(x)\forall x\in\mathbb{R}$, at a fixed initial time $t_0$.
In particular, a special solution is given by the kernel or propagator, defined by the initial condition  $W(x,t_0)=\delta(x-y)$, where $y$ is an arbitrary initial value. It means that the initial value is deterministically known. This solution is referred to as $W(x,t)=K(x,t;y,t_0)$ and it can be obtained by the partial differential problem
\begin{eqnarray}
    \disp \frac{\partial K}{\partial t} &=& -fK-\ell\frac{\partial K}{\partial x}+g^2\frac{\partial^2K}{\partial x^2}\\
    K(x,t_0;y,t_0) &=& \delta(x-y).
\end{eqnarray}
With a little abuse of language, we can say that $K$ is the probability density of observing the value $x$ at time $t$, given the value $y$ at time $t_0$. 

The propagator is important since any initial condition, such as  $W(x,t_0)=W_0(x)\forall x\in\mathbb{R}$, can be handled by the relation
\begin{equation}
    W(x,t)=\int_{-\infty}^{+\infty}K(x,t;y,t_0)W_0(y)\d y,
\end{equation}
as a consequence of the linearity of the Fokker-Planck equation.

The objective of this work is to find a good approximation of the propagator $K$ for short times, and this will be organized
as follows. The short-time approximation obviously contains the singularity induced by the initial condition represented by the delta function $\delta(x-y)$. 
It means that we look for $K_0$ such that $K(x,t;y,t_0)\underset{t\to t_0}{\sim}K_0(x,t;y,t_0)$. 
Successively, for larger values of time, we can introduce a correction function $F$ such that $K(x,t;y,t_0)=K_0(x,t;y,t_0)F(x,t;y,t_0)$. 
Since the effects of the initial singularity are all contained in $K_0$, it follows that $F$ is regular. We will obtain a partial differential equation describing the behavior of $F$.
Moreover, based on the regularity of $F$, we will introduce the Taylor development $F(x,t;y,t_0)=\disp\sum_{n=0}^{+\infty}F_n(x;y)(t-t_0)^n$ and we obtain the coefficient $F_{n+1}$ recursively, i.e. as function of preceding coefficients $F_k, \forall k$ such that $0\leq k\leq n$.
Further simplifications are introduced in the case where the functions $h$ and $g$ are time-independent. 

\subsection{Formal development of the theory}
\label{secformal}

We describe here in detail the formal development of the theory, which includes the determination of the singular short-time part of the propagator and its regular short-time part. This regular  perturbation is developed in series of powers over time, and the coefficients are described by differential equations that are solved explicitly. 

{\it The singular short-time part of the propagator.}
In order to obtain the explicit form of $K_0$ associated to the Fokker-Planck equation stated in Eq. (\ref{fp}), we firstly introduce the basic stochastic differential equation 
\begin{equation}
    \frac{\d x}{\d t}=a\xi(t),
\end{equation}
where $a\in\mathbb{R}$. This represents the simplest Ornstein-Uhlenbeck process \cite{uhlenbeck1930,wang1945}, and the corresponding Fokker-Planck equation is given by the diffusion or heat equation
\begin{equation}
    \frac{\partial K}{\partial t}=a^2\frac{\partial^2K}{\partial x^2}.
\end{equation}
The propagator for this equation is well-known and can be written as follows \cite{risken1989,gardiner2009}
\begin{equation}
    K(x,t;y,t_0)=\frac{1}{\sqrt{4\pi a^2(t-t_0)}}\exp\left[-\frac{(x-y)^2}{4a^2(t-t_0)}\right],
    \label{eqkernelheat}
\end{equation}
when it is determined from the initial condition $K(x,t_0;y,t_0) = \delta(x-y)$. 
This kernel represents the exact solution and perfectly describes the singularity for $t\rightarrow t_0$.

In the heat equation just solved, $a$ is a constant. In contrast, the diffusive term in the more general Fokker-Planck equation given in Eq. (\ref{fp}) arbitrarily depends on space and time. 
Hence, the idea is to consider the following short-time behavior of Eq. (\ref{fp}) 
\begin{equation}
    K_0(x,t;y,t_0)=\frac{C}{\sqrt{4\pi (t-t_0)}}\exp\left[-\frac{(\omega(x)-\omega(y))^2}{4(t-t_0)}\right].
\end{equation}
where $\omega(x)$ is a regular function able to take into account the heterogeneous diffusion. 

The introduction of a function $\omega$, for now arbitrary, is useful to take into account the possible non-Gaussian behavior induced by the arbitrary form of $g(x,t)$. The validity of this proposed form of the short-time propagator is confirmed \textit{a posteriori}, by substitution.

The constant $C$ has been introduced to normalize the propagator and is determined as follows,  by considering the following representation of the delta function
\begin{equation}
\lim_{\varepsilon \rightarrow 0}\frac{1}{\sqrt{\pi \varepsilon}}\exp\left(-\frac{z^2}{\varepsilon}\right)=\delta(z).
\label{limdelta}
\end{equation}  
To this aim, we study the expression $\delta(\omega(x)-\omega(y))$ for an arbitrary $\omega(x)$. Of course, if $x\neq y$ we have $\delta(\omega(x)-\omega(y))=0$. 
Hence, we need to evaluate the integral $\int_{-\infty}^{+\infty}\delta(\omega(x)-\omega(y))\d x$. 
To do this, we use the change of variable $z=\omega(x)$, and we introduce the inverse function $x=\Omega(z)$. 
We suppose that the function $\omega$ is always increasing, as we will be able to confirm later on.
By definition, $\omega(\Omega(z))=z$ and $\Omega(\omega(x))=x$.
Then, we get $\d z=\omega'(x)\d x$ and $\d x= \Omega'(z)\d z$, with $\omega'(x)\ge 0$ and $\Omega'(z) \ge 0$.
We can write
\begin{eqnarray}
\nonumber
    \int_{-\infty}^{+\infty}\delta(\omega(x)-\omega(y))\d x&=&\int_{-\infty}^{+\infty}\delta(z-\omega(y))\Omega'(z)\d z\\
    &=&\Omega'(\omega(y))=\frac{1}{\omega'(y)}.
\end{eqnarray}
Indeed, from $\Omega(\omega(x))=x$, we obtain by derivation $\omega'(x)\Omega'(\omega(x))=1$ and hence $\Omega'(\omega(x))=1/\omega'(x)$.
We finally proved that
\begin{equation}
\delta(\omega(x)-\omega(y))=\disp\frac{1}{\omega'(y)}\delta(x-y).
\end{equation}
This is useful to obtain $C=\omega'(y)$, yielding the normalized expression for the propagator
\begin{equation}
    K_0(x,t;y,t_0)=\frac{\omega'(y)}{\sqrt{4\pi (t-t_0)}}\exp\left[-\frac{(\omega(x)-\omega(y))^2}{4(t-t_0)}\right].
    \label{k0}
\end{equation}
Indeed, by using the classical limiting expression  in Eq. (\ref{limdelta}), we deduce that
\begin{eqnarray}
K_0(x,t;y,t_0)\underset{t \rightarrow t_0}{\longrightarrow} \omega'(y)\delta(\omega(x)-\omega(y))=\delta(x-y),
\end{eqnarray}
confirming the correct short time behavior of Eq. (\ref{k0}).
The function $\omega(x)$ is  useful to accommodate the complexity introduced by the space and time dependence of $f$, $\ell$, and in particular $g$.  
To obtain the explicit time-space dependence, we firstly substitute Eq. (\ref{k0}) into the Fokker-Planck equation. 
Hence, we need the following derivatives:
\begin{eqnarray}
        \frac{\partial K_0}{\partial x} &=& -K_0\frac{\omega(x)-\omega(y)}{2(t-t_0)}\omega'(x),\\
             \frac{\partial^2K_0}{\partial x^2} &=& K_0\left[\frac{\omega(x)-\omega(y)}{2(t-t_0)}\omega'(x)\right]^2\\
           \nonumber
        &&-\frac{K_0}{2(t-t_0)}\left[\omega'(x)^2+(\omega(x)-\omega(y))\omega''(x)\right],\\
        \frac{\partial K_0}{\partial t} &=& -\frac{K_0}{2(t-t_0)}+K_0\frac{(\omega(x)-\omega(y))^2}{4(t-t_0)^2}.
\end{eqnarray}
In Eq. (\ref{fp}), the leading terms for $t\rightarrow t_0$ are those proportional to $1/(t-t_0)^2$. 
We see that there is a leading term in $\frac{\partial^2K_0}{\partial x^2}$, and another one in $ \frac{\partial K_0}{\partial t}$. 
The balance of these two terms leads to
\begin{equation}
    K_0\frac{(\omega(x)-\omega(y))^2}{4(t-t_0)^2}=g(x,t_0)^2K_0\left[\frac{\omega(x)-\omega(y)}{2(t-t_0)}\omega'(x)\right]^2,
\end{equation}
and then $\omega'(x)^2g^2(x,t_0)=1$.
Since $g(x,t_0)\ge 0$, we find $\omega'(x)=1/ g(x,t_0)\ge 0 $, and therefore
\begin{eqnarray}
\omega(x)-\omega(y)=\int_y^x\frac{\d\eta}{ g(\eta,t_0) }\triangleq\mathcal{D}(x,y,t_0),
\label{LLL}
\end{eqnarray}
where we defined the quantity $\mathcal{D}(x,y,t_0)$ for use in following developments.
This calculation confirms that $\omega'(x)\ge 0$, as assumed in previous developments.
To conclude, the short time propagator is given by
\begin{eqnarray}
\nonumber
    K_0(x,t;y,t_0)&=&\frac{\exp\left[-\frac{1}{4(t-t_0)}\left(\int_y^x\frac{\d\eta}{ g(\eta,t_0) }\right)^2\right]}{ g(y,t_0) \sqrt{4\pi(t-t_0)}}\\
    &=&\frac{\exp\left[-\frac{1}{4(t-t_0)}\mathcal{D}^2(x,y,t_0)\right]}{ g(y,t_0) \sqrt{4\pi(t-t_0)}},
    \label{short}
\end{eqnarray}
and it is only influenced by the diffusion coefficient $g$ for the
initial time $t=t_0$.
It is important to remark that the integral in Eq. (\ref{LLL}) corresponds to the geodetic distance in a multi-dimensional setting, as discussed in Ref. \cite{bilal2020}. Interestingly, the geodetic distance $\mathcal{D}$ can be written in closed form, as in Eq. (\ref{LLL}), only for the one-dimensional case considered here.  

{\it The regular short-time part of the propagator.}
In order to improve the short-time representation of the propagator, we can consider the following solution
\begin{equation}
    K(x,t;y,t_0)=K_0(x,t;y,t_0)F(x,t;y,t_0),
    \label{kof}
\end{equation}
where $F$  is for the moment unknown. 
Of course, the choice to consider the multiplicative correction is arbitrary, and an additive correction could have been used. We kept the first choice because it involves analytically feasible calculations and exact solutions as discussed below and in the recent literature \cite{nistor2010,nistor2011,bilal2020}. 

An equation for the quantity $F$ can be found by substituting Eq. (\ref{kof}) into Eq. (\ref{fp}), and by considering the expression just obtained for $K_0$.
First of all, we introduce the notations $g_0(x)=g(x,t_0)$ and $g_0'(x)=\partial g(x,t_0)/\partial x$, and we calculate the following partial derivatives
\begin{eqnarray}
        \frac{\partial K}{\partial x}&=&-\frac{K_0}{2(t-t_0)}\frac{\mathcal{D}}{g_0}F+K_0\frac{\partial F}{\partial x}\\
        \frac{\partial K}{\partial t}&=&-\frac{K_0F}{2(t-t_0)}+\frac{\mathcal{D}^2K_0F}{4(t-t_0)^2}+K_0\frac{\partial F}{\partial t}\\
        \nonumber
        \frac{\partial^2 K}{\partial x^2}&=&\frac{K_0}{4(t-t_0)^2}\frac{\mathcal{D}^2}{g_0^2}F-\frac{K_0}{2(t-t_0)}\frac{F}{g_0^2}+\frac{K_0}{2(t-t_0)}\frac{\mathcal{D}g_0'}{g_0^2}F\\
        &&-\frac{K_0}{(t-t_0)}\frac{\mathcal{D}}{g_0}\frac{\partial F}{\partial x}+K_0\frac{\partial^2F}{\partial x^2}.
\end{eqnarray}
By using these expressions in Eq. (\ref{fp}), we elaborate an equation for the function $F$
\begin{eqnarray}
\nonumber
    \frac{\partial F}{\partial t}&=&-fF-\ell\frac{\partial F}{\partial x}+g^2\frac{\partial^2F}{\partial x^2}-\frac{1}{2(t-t_0)}F\left(\frac{g^2}{g_0^2}-1\right)\\
    \label{eqf}
    &&+\frac{1}{4(t-t_0)^2}\mathcal{D}^2F\left(\frac{g^2}{g_0^2}-1\right)\\
    \nonumber
    &&+\frac{\mathcal{D}}{2(t-t_0)g_0^2}\left(\ell Fg_0+g^2g_0'F-2g^2g_0\frac{\partial F}{\partial x}\right).
\end{eqnarray}
Although this result appears in a more complex form than the initial Fokker-Planck equation, the function $F$ always has a regular behavior for each time $t\ge t_0$. In such an equation, it must be taken care that $g(x,t)$ is different from $g_0(x)=g(x,t_0)$. We underline two specific points:

(i) there is only one initial condition to fully determine $F$, which is
\begin{equation}
    F(y,t_0;y,t_0)=1.
    \label{ic}
\end{equation}
It means that the function $F$  must not alter the unit integral of the delta function for $t\rightarrow t_0$ and, therefore, it must be equal to one only for $x=y$.
 
 (ii) since the singular behavior for $t\rightarrow t_0$ is fully described by $K_0$, the function $F$ is regular everywhere and can be developed in a Taylor series with respect to $t-t_0$.
 
According to this second point, since all the functions $F$, $f$, $\ell$, and $g^2$, involved in Eq. (\ref{eqf}) are regular, we can introduce the following analytic developments
\begin{eqnarray}
        \label{serF}
        \disp F(x,t;y,t_0)&=&\sum_{n=0}^{+\infty}F_n(x;y)(t-t_0)^n,\\
        \label{serf}
        \disp f(x,t)&=&\sum_{n=0}^{+\infty}f_n(x)(t-t_0)^n,\\
        \label{serl}
        \disp \ell(x,t)&=&\sum_{n=0}^{+\infty}\ell_n(x)(t-t_0)^n,\\     \label{serg}
        \disp g^2(x,t)&=&\sum_{n=0}^{+\infty}g_n^2(x)(t-t_0)^n.
\end{eqnarray}
The initial condition stated in Eq. (\ref{ic}) simplifies to
\begin{equation}
\label{cond}
    F_0(y;y)=1.
\end{equation}
We will discuss later the conditions for $F_n$, $n\ge 1$, which need to be automatically identified by the procedure we are developing since there are no other \textit{a priori} conditions to be imposed.
We suppose from now on that all terms $f_n$, $\ell_n$, and $g_n^2$ are known regular functions of $x\in\mathbb{R}$, $n\ge 0$.

Before starting the substitution of Eqs. (\ref{serF}), (\ref{serf}), (\ref{serl}), and (\ref{serg}) into Eq. (\ref{eqf}), we observe that
\begin{equation}
 \frac{g^2}{g_0^2}-1=\sum_{n=1}^{+\infty}\frac{g_n^2}{g_0^2}(t-t_0)^n,
\end{equation}
which means that ${g^2}/{g_0^2}-1$ is divisible by $t-t_0$.
Therefore, the last term in the first line of Eq. (\ref{eqf}) is regular, i.e. without the singularity $1/(t-t_0)$, and the term in the second line exhibits a singularity of the type $1/(t-t_0)$ instead of type $1/(t-t_0)^2$, as it might seem at a first glance.
Thus, the term in the second line together with the terms in the third line share the same singularity of the type $1/(t-t_0)$. Importantly, we will see that the balance equation for these terms will yield the first coefficient $F_0(x;y)$.
That said, we can substitute all Taylor developments into the main equation and obtain, after long calculations, the relation
\begin{eqnarray}
\nonumber
        \sum_{n=0}^{+\infty}F_{n+1}(n+1)(t-t_0)^{n}&=&\sum_{n=0}^{+\infty}\mathcal{G}_n(t-t_0)^n\\
        \nonumber
        &+&\frac{\mathcal{D}^2g_1^2}{4(t-t_0)g_0^2}F_0+\frac{\mathcal{D}\ell_0}{2(t-t_0)g_0}F_0\\
        \label{big}
        &+&\frac{\mathcal{D}g_0'}{2(t-t_0)}F_0-\frac{\mathcal{D}g_0}{(t-t_0)}\frac{\partial F_0}{\partial x},
        \label{super}
\end{eqnarray}
where
\begin{eqnarray}
\nonumber
        \mathcal{G}_n&=&\sum_{k=0}^n\left[-f_kF_{n-k}-\ell_k\frac{\partial F_{n-k}}{\partial x}+g_k^2\frac{\partial^2F_{n-k}}{\partial x^2}-\frac{g_{k+1}^2}{2g_0^2}F_{n-k}\right]\\
        \nonumber
        &&+\sum_{k=0}^{n+1}\left[\frac{\mathcal{D}^2}{4}\frac{g_{k+1}^2}{g_0^2}F_{n+1-k}+\frac{\mathcal{D}}{2g_0}\ell_kF_{n+1-k}\right.\\
        &&\left.+\frac{\mathcal{D}g_0'}{2g_0^2}g_k^2F_{n+1-k}-\frac{\mathcal{D}}{g_0}g_k^2\frac{\partial F_{n+1-k}}{\partial x}\right].
\end{eqnarray}
The last four terms of Eq. (\ref{super}), with the singularity ${1}/{(t-t_0)}$, give the following equation for $F_0$
\begin{equation}
    \frac{\partial F_0}{\partial x}=\left(\frac{\mathcal{D}g_1^2}{4g_0^3}+\frac{\ell_0}{2g_0^2}+\frac{g_0'}{2g_0}\right)F_0,
\end{equation}
where $F_0=F_0(x;y)$, with $F_0(y;y)=1$. Hence, the solution is easily obtained as
\begin{eqnarray}
\nonumber
    F_0(x;y)&=&\exp\left[\frac{1}{4}\int_y^x\frac{\mathcal{D}(\eta,y,t_0)g_1^2(\eta)}{g_0^3(\eta)}\d\eta\right.\\
    \label{f0}
    &&\left.+\frac{1}{2}\int_y^x\frac{\ell_0(\eta)}{g_0^2(\eta)}\d\eta+\frac{1}{2}\int_y^x\frac{g_0'(\eta)}{g_0(\eta)}\d\eta\right].
\end{eqnarray}
We therefore obtained a closed form expression for the first correction term $F_0$. 
The higher-order corrections can be found by equaling to zero the sum of terms having the same power $(t-t_0)^n$, $n\ge 0$, in Eq. (\ref{big}). 
This results in a differential equation for $F_{n+1}$, $n\ge 0$, that depends on all the preceding terms $F_0$,...,$F_n$. It reads
\begin{eqnarray}
\nonumber
        &&\mathcal{D}g_0\frac{\partial F_{n+1}}{\partial x}+F_{n+1}\left(n+1-\frac{\mathcal{D}^2g_1^2}{4g_0^2}-\frac{\mathcal{D}\ell_0}{2g_0}-\frac{\mathcal{D}g_0'}{2}\right)        \\
        \nonumber
        &=&\sum_{k=0}^n\bigg(-f_kF_{n-k}-\ell_k\frac{\partial F_{n-k}}{\partial x}+g_k^2\frac{\partial^2F_{n-k}}{\partial x^2}-\frac{g_{k+1}^2}{2g_0^2}F_{n-k}\\
        \nonumber
        &&+\frac{\mathcal{D}^2g_{k+2}^2}{4g_0^2}F_{n-k}+\frac{\mathcal{D}\ell_{k+1}}{2g_0}F_{n-k}\\
        \label{fnpu}
        &&+\frac{\mathcal{D}g_0'g_{k+1}^2}{2g_0^2}F_{n-k}-\frac{\mathcal{D}g_{k+1}^2}{g_0}\frac{\partial F_{n-k}}{\partial x}\bigg).
\end{eqnarray}
An important point concerns the initial condition to be used for this differential equation. 
The original condition $F(y,t_0;y,t_0)=1$, simplified to $F_0(y;y)=1$, has been already used in Eqs.(\ref{cond}) and (\ref{f0}). 
So, there are no other conditions from the original problem that we are studying. 
However, we easily see that the term with ${\partial F_{n+1}}/{\partial x}$ in Eq. (\ref{fnpu}) is proportional to $\mathcal{D}(x,y,t_0)$.
It means that if we consider $x=y$ in Eq. (\ref{fnpu}) we obtain  a relation fixing $F_{n+1}(y;y)$, since $\mathcal{D}(y,y,t_0)=0$, see Eq. (\ref{LLL}).
The initial condition is therefore
\begin{eqnarray}
\nonumber
    F_{n+1}(y;y)&=&\frac{1}{n+1}\sum_{k=0}^n\bigg(-f_kF_{n-k}-\ell_k\frac{\partial F_{n-k}}{\partial x}\\
    && +g_k^2\frac{\partial^2F_{n-k}}{\partial x^2}-\frac{g_{k+1}^2}{2g_0^2}F_{n-k}\bigg).
\end{eqnarray}
From a conceptual point of view, the problem we studied was traced back to the solution of a hierarchy of differential problems, i.e. an infinite sequence of differential equations whose initial conditions are known.  
Nevertheless, although these equations are of the first order, their form is quite complex and needs to be simplified as described below.

\subsection{The formal solution of the hierarchy of differential equations}
\label{formal}

In order to simplify our results from section \ref{secformal}, we define
\begin{eqnarray}
        \Phi(x,y)&\equiv&\frac{\mathcal{D}g_1^2}{4g_0^3}+\frac{\ell_0}{2g_0^2}+\frac{g_0'}{2g_0},\\
        \nonumber
        \Psi_n(x,y)&\equiv&\sum_{k=0}^{n-1}\left[-f_kF_{n-1-k}-\ell_k\frac{\partial F_{n-1-k}}{\partial x}+g_k^2\frac{\partial^2F_{n-1-k}}{\partial x^2}\right.\\
        \nonumber
        &&-\frac{g_{k+1}^2}{2g_0^2}F_{n-1-k}+\frac{\mathcal{D}^2g_{k+2}^2}{4g_0^2}F_{n-1-k}+\frac{\mathcal{D}\ell_{k+1}}{2g_0}F_{n-1-k}\\
        &&\left.+\frac{\mathcal{D}g_0'g_{k+1}^2}{2g_0^2}F_{n-1-k}-\frac{\mathcal{D}g_{k+1}^2}{g_0}\frac{\partial F_{n-1-k}}{\partial x}\right],
\end{eqnarray}
and the differential equations, with corresponding initial conditions, can be rewritten as 
\begin{eqnarray}
    \frac{\partial F_0}{\partial x}-\Phi F_0&=&0,\\
    F_0(y,y)&=&1,\\
    \label{eqfn}
    \frac{\partial F_n}{\partial x}+\left(n\frac{\mathcal{D}'}{\mathcal{D}}-\Phi\right)F_n&=&\frac{\mathcal{D}'}{\mathcal{D}}\Psi_n,\\
    \label{condfn}
    F_n(y;y)&=&\frac{1}{n}\Psi_n(y,y),
\end{eqnarray}
where we considered $n\ge 1$, and we used the fact that $g_0(x)=1/\mathcal{D}'(x,y,t_0)$, with $\mathcal{D}'=\partial\mathcal{D}/\partial x$. Since the quantity $\mathcal{D}$ depends on $y$, while  $\mathcal{D}'$ does not depend on $y$, from now on, we use consistently the notation $\mathcal{D}=\mathcal{D}(x,y,t_0)$ and $\mathcal{D}'=\mathcal{D}'(x,t_0)$.
Clearly, the solution of the differential equation for $F_0$ is
\begin{equation}
    F_0(x;y)=\exp\left[\int_y^x\Phi(\eta,y)d\eta \right],
    \label{f0bis}
\end{equation}
which exactly corresponds to Eq. (\ref{f0}).
For solving Eq. (\ref{eqfn}), we start by considering the homogeneous equation ${\partial F_n^h}/{\partial x}+\left(n{\mathcal{D}'}/{\mathcal{D}}-\Phi\right)F_n^h=0$, and we obtain the solution as
\begin{eqnarray}
        \nonumber
        F_n^h(x;y)&=&\exp\left[-n\int_z^x\frac{\mathcal{D}'(\eta,t_0)}{\mathcal{D}(\eta,y,t_0)}\d\eta+\int_z^x\Phi(\eta,y)\d\eta\right]\\
        \nonumber
        &=&\exp\left[-n\ln\left(\frac{\mathcal{D}(x,y,t_0)}{\mathcal{D}(z,y,t_0)}\right)+\int_z^x\Phi(\eta,y)\d\eta\right]\\
        &=&\left(\frac{\mathcal{D}(z,y,t_0)}{\mathcal{D}(x,y,t_0)}\right)^n\exp\left[\int_z^x\Phi(\eta,y)\d\eta\right],
\end{eqnarray}
where $\ln(\zeta)$ represents the natural logarithm of $\zeta$. Moreover, $z$ takes an arbitrary value that could be fixed by knowing the initial condition of $F_n^h$. 
For now, let's leave $z$ free and look for the solution of the non-homogeneous equation by adopting the method of variation of parameters (Lagrange’s method). 
We can in fact write a particular solution in the form
\begin{equation}
\label{test}
    F_n^p(x;y)=C(x)\left(\frac{\mathcal{D}(z,y,t_0)}{\mathcal{D}(x,y,t_0)}\right)^n\exp\left[\int_z^x\Phi(\eta,y)\d\eta\right],
\end{equation}
where  $C(x)$ is an unknown regular function, which can be determined as follows. 
We can substitute Eq. (\ref{test}) into Eq. (\ref{eqfn}) to get
\begin{eqnarray}
\nonumber
    &&C'(x)\left(\frac{\mathcal{D}(z,y,t_0)}{\mathcal{D}(x,y,t_0)}\right)^n\exp\left[\int_z^x\Phi(\eta,y)\d\eta\right]\\
    &&\,\,\,\,\,\,\,\,\,\,\,\,\,\,\,\,\,\,=\frac{\mathcal{D}'(x,t_0)}{\mathcal{D}(x,y,t_0)}\Psi_n(x,y).
\end{eqnarray}
Hence, we find the following explicit expression for $C(x)$
\begin{eqnarray}
\nonumber
    C(x)&=&\frac{1}{\mathcal{D}(z,y,t_0)^n}\int_y^x\mathcal{D}'(\eta,t_0)\mathcal{D}(\eta,y,t_0)^{n-1}\Psi_n(\eta,y)\\
    &&\times\exp\left[-\int_z^{\eta}\Phi(\chi,y)\d\chi\right]\d\eta.
\end{eqnarray}
A particular solution of Eq. (\ref{eqfn}) is therefore
\begin{eqnarray}
\nonumber
        &&F_n^p(x;y)=\frac{1}{\mathcal{D}(x,y,t_0)^n}\int_y^x\mathcal{D}'(\eta,t_0)\mathcal{D}(\eta,y,t_0)^{n-1}\Psi_n(\eta,y)\\
        \nonumber
    &&\times\exp\left[-\int_z^{\eta}\Phi(\chi,y)\d\chi\right]\d\eta\exp\left[\int_z^x\Phi(\eta,y)\d\eta\right],\\
\end{eqnarray}
which is valid for any value of $z$.
In general, following Lagrange theory, the complete solution of Eq. (\ref{eqfn}) is given by the sum of the solution of the homogeneous equation and the particular solution. 
In this specific case, the procedure can be simplified by trying to consider $z=y$ in previous solutions. 
So doing, the solution of the homogeneous equation is zero since $\mathcal{D}(y,y,t_0)=0$, and the general solution becomes coincident with the particular solution with $z=y$ ($n\ge 1$)
\begin{eqnarray}
\nonumber
        &&F_n(x;y)=\frac{1}{\mathcal{D}(x,y,t_0)^n}\int_y^x\mathcal{D}'(\eta,t_0)\mathcal{D}(\eta,y,t_0)^{n-1}\\
        \nonumber
    &&\times\Psi_n(\eta,y)\exp\left[-\int_y^{\eta}\Phi(\chi,y)\d\chi\right]\d\eta\exp\left[\int_y^x\Phi(\eta,y)\d\eta\right].\\
    \label{finfn}
\end{eqnarray}
We can verify that this solution perfectly satisfies the initial condition $F_n(y;y)=\Psi_n(y,y)/n$. 
To this aim, we consider two regular functions $\phi(x)$ and $a(x)$ such that $\phi(y)=0$, and we use the result
\begin{equation}
\lim_{x\to y}\frac{1}{\phi(x)}\int_y^xa(\eta)\phi'(\eta)\d\eta=a(y),
\end{equation}
which can be directly proved, for instance, by means of L'Hôpital's rule.
We can assume $\phi(x)=\mathcal{D}(x,y,t_0)^n$ so that $\phi'(x)=n\mathcal{D}'(x,t_0)\mathcal{D}(x,y,t_0)^{n-1}$, and $a(x)=\frac{1}{n}\Psi_n(x,y)\exp\left(-\int_y^x\Phi(\chi,y)\d\chi\right)$. 
Then, we find that $\lim_{x\rightarrow y} F_n(x;y)=\Psi_n(y,y)/n$, as expected. 

To conclude, Eqs.(\ref{f0bis}) and (\ref{finfn}) are the formal solutions of the hierarchy of differential equations for the coefficients $F_n$. 
Thus we have at our disposal the complete mathematical form of the propagator $K=K_0F$ of the Fokker-Planck equation, consisting in the product of Eq. (\ref{short}) and Eq. (\ref{serF}), where the coefficients just found are to be used.

A particular but very important case concerns the time-independent stochastic differential equation
\begin{equation}
    \frac{\d x}{\d t}=h(x)+g(x)\xi(t),
    \label{lanti}
\end{equation}
In this case the functions $f$ and $\ell$ remains unchanged with respect to Eqs.(\ref{effe}) and (\ref{elle}), but their Taylor developments are greatly simplified.  Indeed, we have $f_0(x)=f(x)$, $\ell_0(x)=\ell(x)$, and $g_0^2(x)=g^2(x)$. Moreover,  for all $n\ge 1$, we have $f_n(x)=0$, $\ell_n(x)=0$, and $g_n^2(x)=0$.
As a result, the functions $\Phi$ and $\Psi_n$ turn out to be simplified as follows   
   \begin{eqnarray}
   \label{phiti}
        \Phi&=&\frac{\ell}{2g^2}+\frac{g'}{2g},\\
        \label{psiti}
        \Psi_n&=&-fF_{n-1}-\ell\frac{\partial F_{n-1}}{\partial x}+g^2\frac{\partial^2F_{n-1}}{\partial x^2}
\end{eqnarray}
Importantly, $\Psi_n$ depends only on the previous coefficient $F_{n-1}$ and not on all coefficients as in the arbitrarily time-varying case. 
So, in this case Eq. (\ref{finfn}) represents a simple recursion between $F_n$  and $F_{n-1}$.
This makes the procedure easily applicable in many practical cases. 
In addition, this approach can be further simplified by defining
   \begin{equation}
   \label{new}
       F_n(x;y)=D_n(x,y)F_0(x;y).
   \end{equation}
      Of course, we have $D_0=1$. 
   As discussed below, the recursion law for the parameters $D_n$ is simpler than the previous one, concerning the parameters $F_n$. 
  To prove this point, we begin by developing the derivatives in Eq. (\ref{psiti}), as follows
\begin{eqnarray}
        \frac{\partial F_k}{\partial x}&=&\left(\frac{\partial D_k}{\partial x}+D_k\Phi\right)F_0\\
        \frac{\partial^2F_k}{\partial x^2}&=&\left(\frac{\partial^2D_k}{\partial x^2}+2\frac{\partial D_k}{\partial x}\Phi+D_k\frac{\partial \Phi}{\partial x}+D_k\Phi^2\right)F_0.
\end{eqnarray}
It is important to remark that both results are proportional to $F_0$.
It means that we can introduce a function $\Lambda_n$, such that $\Psi_n=\Lambda_nF_0$, which can be eventually written as
\begin{eqnarray}
\label{lambda}
        \Lambda_n&=&-fD_{n-1}-\ell\left(\frac{\partial D_{n-1}}{\partial x}+D_{n-1}\Phi\right)\\
        \nonumber
        &&+g^2\bigg(\frac{\partial^2 D_{n-1}}{\partial x^2}+2\frac{\partial D_{n-1}}{\partial x}\Phi+D_{n-1}\frac{\partial \Phi}{\partial x}+D_{n-1}\Phi^2\bigg).
\end{eqnarray}
In this time-independent situation, the function $\mathcal{D}$ can be redefined as $\mathcal{D}(x,y)=\int_y^x{\d\eta}/{ g(\eta) }$ and, therefore, $\mathcal{D}'(x)=1/ g(x) $. Hence, Eq. (\ref{finfn}) takes the form
\begin{eqnarray}
\nonumber
        F_n(x;y)&=&\int_y^x\frac{\mathcal{D}'(\eta)\mathcal{D}(\eta,y)^{n-1}\Psi_n(\eta,y)}{\mathcal{D}(x,y)^nF_0(\eta,y)}\d\eta F_0(x;y)\\
            \nonumber
    &=&\int_y^x\frac{\mathcal{D}'(\eta)\mathcal{D}(\eta,y)^{n-1}\Lambda_n(\eta,y)}{\mathcal{D}(x,y)^n}\d\eta F_0(x;y),\\
\end{eqnarray}
or, equivalently
\begin{equation}
\label{dn}
    D_n(x,y)=\frac{1}{\mathcal{D}(x,y)^n}\int_y^x\mathcal{D}'(\eta)\mathcal{D}(\eta,y)^{n-1}\Lambda_n(\eta,y)\d\eta.
\end{equation}
Finally, the combination of Eqs.(\ref{lambda}) and (\ref{dn}) represents a recursive procedure that gives all the parameters $D_n$, which, in turn, can be used to obtain all the parameters $F_n$ through Eq. (\ref{new}).
This approach can be easily applied to study the behavior of stochastic differential equations, as in Eq. (\ref{lanti}), with arbitrary drift and heterogeneous diffusion. 

In the next section, we will turn to applications of our formalism.
In particular, we demonstrate the determination of the coefficients $D_n$, and the convergence of the series in Eq. (\ref{serF}) for various stochastic processes.
\\

\section{Application to specific stochastic processes}
\label{secapp}

We now turn to the application of our formal developments to specific examples. We begin with the simplest of all cases, a Gaussian process, which has numerous applications in many scientific fields. The second is the geometric Brownian process, a stochastic
process most often associated with problems in finance. It is worth noting that for these first two examples we could compute the full propagator exactly, against which the form of short-term propagator can be easily compared. 
The third and fourth are two case studies taken from biophysics, for which we do not know a priori the full propagator: (i) a model stochastic process
for a molecular motor, and (ii) a stochastic process occurring in parasite motility, based on an exponential heterogeneous diffusion coefficient. 

\subsection{Gaussian processes}

This simple case describes a system with constant drift and diffusion terms. We consider therefore two real parameters $H_0$ and $G_0>0$ and the stochastic differential equation
\begin{equation}
    \frac{\d x}{\d t}=-H_0+G_0\xi(t).
\end{equation}
Of course, the propagator for this problem is well-known, in fact a generalization of Eq. \ref{eqkernelheat} that can be written as \cite{risken1989,gardiner2009}
\begin{equation}
\label{solgauss}
    K(x,t;y,t_0)=\frac{\exp\left[-\frac{(x-y+H_0(t-t_0))^2}{4G_0^2(t-t_0)}\right]}{\sqrt{4\pi G_0^2(t-t_0)}}.
\end{equation}
Within our formalism we thus have for Eqs. (\ref{effe}) and (\ref{elle}) $f(x)=0$, $\ell(x)=-H_0$, and $g^2(x)=G_0^2$. Moreover, $\mathcal{D}(x,y)=(x-y)/G_0$. Therefore, we can obtain the 
short-time propagator $K_0$ in the form
\begin{equation}
    K_0(x,t;y,t_0)=\frac{1}{\sqrt{4\pi G_0^2(t-t_0)}}\exp\left[-\frac{(x-y)^2}{4G_0^2(t-t_0)}\right].
\end{equation}
It can be immediately seen that this expression, valid for short times, does not coincide with $K$, Eq. (\ref{solgauss}). Therefore, the correction introduced by the 
function $F$  is essential in this case to obtain the correct propagator. 

To determine this correction, we evaluate $\Phi=-\disp\frac{H_0}{2G_0^2}$, and we write $F_0$ as 
\begin{equation}
    F_0(x;y)=\exp\left[\int_y^x\Phi(\eta)\d\eta\right]=\exp\left[-\frac{H_0}{2G_0^2}(x-y)\right].
\end{equation}

\begin{figure}[htb]
\centering
\includegraphics[scale=0.60]{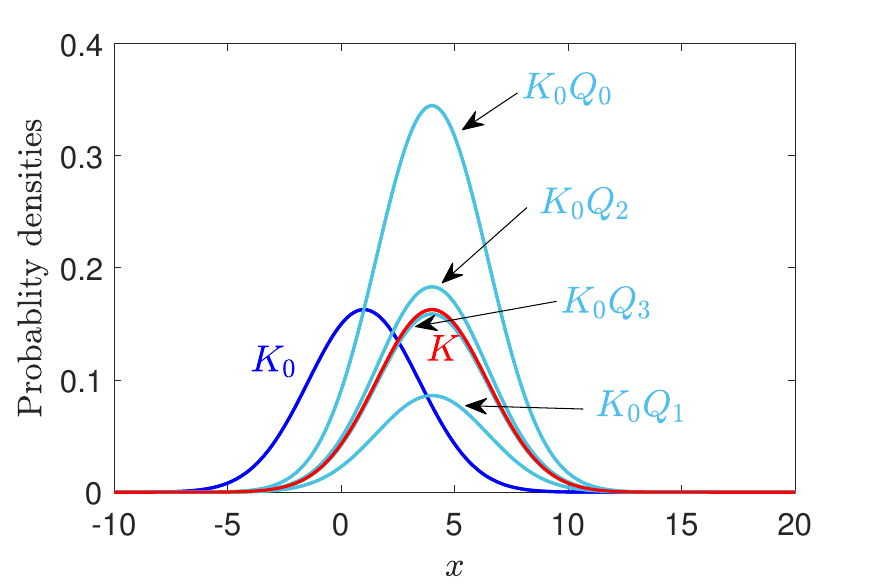}
\caption{\label{exgauss} Demonstration of the convergence of the expansion for a Gaussian process. We plot the short-time propagator $K_0$, the full propagator $K$, and low-order approximations $K_0Q_0$, $K_0Q_1$, $K_0Q_2$, and $K_0Q_3$, to show the convergence $K_0Q_m\to K $ when $m\to \infty$. We adopted the parameter values $H_0=-1$, $G_0=1$, $y=1$, $t_0=0$, $t=3$ (in arbitrary units).}
\end{figure}
We can prove by mathematical induction that 
\begin{equation}
\label{dngauss}
   D_n=\disp\frac{(-1)^nH_0^{2n}}{4^n n!G_0^{2n}}. 
\end{equation}
This expression was found by explicitly applying the recursive procedure - see Eqs.(\ref{lambda}) and (\ref{dn}) - to the first terms and looking for a general formula, which must now be rigorously proved. To begin with, we note that the formula gives the correct value $D_0=1$, for $n=0$. 
Now, we can calculate $D_{n+1}$ with the procedure of Section \ref{formal} to verify that the assumption made on $D_n$ is correct. 
To this aim, we firstly determine $\Lambda_{n+1}$ through Eq. (\ref{lambda}) as follows
\begin{eqnarray}
    \Lambda_{n+1}=\frac{(-1)^{n+1}H_0^{2(n+1)}}{4^{n+1}n!G_0^{2(n+1)}}.
\end{eqnarray}
Then, we can determine $D_{n+1}$ by means of Eq. (\ref{dn})
\begin{eqnarray}
\nonumber
        D_{n+1}&=&\int_y^x\frac{\mathcal{D}'(\eta)\mathcal{D}(\eta,y)^{n}}{\mathcal{D}(x,y)^{n+1}}\frac{(-1)^{n+1}H_0^{2(n+1)}}{4^{n+1}n!G_0^{2(n+1)}}\d\eta\\
        \nonumber
        &=&\int_y^x\frac{(\eta-y)^n}{(x-y)^{n+1}}\frac{(-1)^{n+1}H_0^{2(n+1)}}{4^{n+1}n!G_0^{2(n+1)}}\d\eta\\
            &=&\frac{(-1)^{n+1}H_0^{2(n+1)}}{4^{n+1}(n+1)!G_0^{2(n+1)}},
\end{eqnarray}
which is consistent with Eq. (\ref{dngauss}) through the replacement $n\to n+1$.
This confirms the validity of the result by mathematical induction. 

This approach, based on an \textit{ansatz} like that in Eq. (\ref{dngauss}), which is then verified by induction on the natural numbers is interesting but can probably be applied to a variety of special cases but not to the general problem. In fact, in the most complicated cases it is difficult to propose an expression for $D_n$ by applying only a few recursions of the procedure. However, it remains useful for various applications discussed below.

We can now explicitly determine $F$  as follows
\begin{eqnarray}
\nonumber
        F(x,t;y,t_0)&=&\sum_{n=0}^{+\infty}F_n(x;y)(t-t_0)^n\\
        \nonumber
        &=&\sum_{n=0}^{+\infty}D_n(x)(t-t_0)^nF_0(x;y)\\
        \nonumber
        &=&\sum_{n=0}^{+\infty}\frac{\left(-\frac{H_0^2(t-t_0)}{4G_0^2}\right)^n}{n!}F_0(x;y)\\
        &=&\exp\left[-\frac{H_0^2(t-t_0)}{4G_0^2}\right]\exp\left[-\frac{H_0}{2G_0^2}(x-y)\right].\,\,\,\,\,\,\,\,\,\,\,\,\,\,\,
        \label{expexp}
\end{eqnarray}
We can finally multiply $K_0$ by $F$, and it is easily seen that we get the correct solution for $K$, given by Eq. (\ref{solgauss}). The series defining the function $F$ is clearly convergent and thus  determines the first exponential function in Eq. (\ref{expexp}).
Although rather simple, this example shows that the coefficients $D_n$ decrease rapidly with $n$ and thus a few terms are often sufficient to have a good approximation on short-time intervals.

An illustration of this behaviour is found in Fig. \ref{exgauss}, where we show $K_0$, $K$, and different approximations given by $K_0Q_m$, where $Q_m=\sum_{n=0}^{m}F_n(x;y)(t-t_0)^n$; $Q_m$ is partial sum of Eq. (\ref{expexp}).
We see that the first approximation $K_0Q_0$ is too crude and in fact is not even properly normalizable. 
On the other hand, the subsequent approximations are automatically normalized to within a negligible error.
To conclude, we finally remark that the results of this first example are independent of $\alpha$ since the noise is additive and not multiplicative. We next turn to a first example with multiplicative noise.

\subsection{Geometric Brownian processes}
\label{appgeo}

Geometric Brownian processes find application in physics, e.g., in the statistical theory of turbulence \cite{fuchs2022,birnir2013}, but they are much more popular in finance, in particular in the modeling of stock option prices \cite{mantegna2000,bouchaud2009,hull2021,Stojkoski2020}.

For this example we will see that we can compute the kernel with our recursive procedure for an arbitrary interpretation of the stochastic integrals (i.e. for any value of $\alpha$).
We consider an exemplary geometric Brownian process described 
by the stochastic differential equation
\begin{equation}
    \frac{\d x}{\d t}=-H_0x+G_0\vert x\vert\xi(t),
\end{equation}
where we considered the absolute value in the diffusion term in order to have  $g(x)\ge 0$, $x\in \mathbb{R}$, that is a key assumption of all our work (considering $G_0>0$). 
We have $h(x)=-H_0x$, $g(x)=G_0\vert x\vert$, and $g'(x)=G_0\sgn(x)$, where $\sgn(\zeta)$ represents the signum function returning the sign of $\zeta$, i.e., $\sgn(\zeta)=+1$ if $\zeta>0$, $\sgn(\zeta)=0$ if $\zeta=0$, and $\sgn(\zeta)=-1$ if $\zeta<0$. 
It easily follows from Eqs.(\ref{effe}), (\ref{elle}), and (\ref{phiti}) that 
\begin{eqnarray}
f(x)&=&-H_0+2(\alpha-1)G_0^2,\\ \ell(x)&=&\left[-H_0+2(\alpha-2)G_0^2\right]x,\\
\Phi(x)&=&\left[-\frac{H_0}{G_0^2}+(2\alpha-3) \right]\frac{1}{2x},
\end{eqnarray}
valid for any $x\in \mathbb{R}$.
An important point concerns the determination of the function $\mathcal{D}(x,y)$, defined as
\begin{eqnarray}
\mathcal{D}(x,y)=\int_y^x\frac{\d\eta}{ g(\eta) }=\frac{1}{G_0}\int_y^x\frac{\d\eta}{ \vert\eta\vert }.
\end{eqnarray}
We know that $\int{\d\eta}/{ \vert\eta\vert }=\sgn(\eta)\ln\vert\eta\vert+C$ (for some real constant $C$), but we must be careful in the use of this indefinite integral.
In fact, if $x$ and $y$ have the same sign (both positive or both negative), there is no problem and the result can be written without doubt as $\int_y^x{\d\eta}/{ \vert\eta\vert }=\sgn(x)\ln\vert x/y \vert=\sgn(y)\ln\vert x/y \vert$.
However, when $\sgn(x)\neq\sgn(y)$, the integration region certainly crosses  the origin and the definite integral is then divergent, i.e. $\int_y^x{\d\eta}/{ \vert\eta\vert }\to\pm\infty$.
To summarize, we have:
\begin{eqnarray}
\mathcal{D}(x,y)=\left\lbrace \begin{array}{cc}
   \frac{1}{G_0}\sgn(x)\ln\vert\ \frac{x}{y} \vert  & \mbox{ if } \sgn(x)=\sgn(y),\\
  \pm\infty   & \mbox{ if } \sgn(x)\neq\sgn(y).
\end{array}\right.
\end{eqnarray}

\begin{figure}[t]
\centering
\includegraphics[scale=0.60]{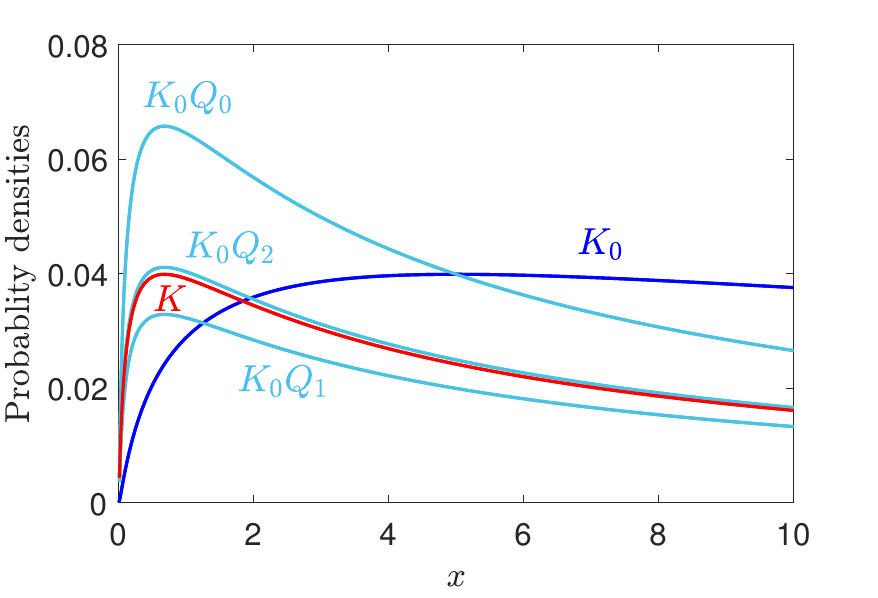}
\caption{\label{exgeo} Convergence of the short-time expansion for a geometric Brownian process. We plot the short-time propagator $K_0$, the full propagator $K$, and some low-order approximations $K_0Q_0$, $K_0Q_1$, and $K_0Q_2$, to illustrate the convergence $K_0Q_m\to K $ when $m\to \infty$. We adopted the parameter values $H_0=-1$, $G_0=1$, $y=5$, $t_0=0$, $t=2$, and $\alpha=1/2$ (in arbitrary units).}
\end{figure}

This complex behavior of the  function $\mathcal{D}$ has important implications for the short-time propagator structure. To begin, we can write 
\begin{equation}
    K_0(x,t;y,t_0)=\frac{1}{ G_0\vert y \vert \sqrt{4\pi(t-t_0)}}\exp\left[-\frac{\mathcal{D}^2(x,y)}{4(t-t_0)}\right],
\end{equation}
and therefore, we obtain
\begin{eqnarray}
    K_0(x,t;y,t_0)&=&\frac{\exp\left[-\frac{\ln^2\vert \frac{x}{y} \vert }{4G_0^2(t-t_0)}\right]}{ G_0\vert y \vert \sqrt{4\pi(t-t_0)}}\mathbf{1}(x),\,\,\,\,\,\,y>0,\\ 
    K_0(x,t;y,t_0)&=&\frac{\exp\left[-\frac{\ln^2\vert \frac{x}{y} \vert }{4G_0^2(t-t_0)}\right]}{ G_0\vert y \vert \sqrt{4\pi(t-t_0)}}\mathbf{1}(-x),\,\,\,\,\,\,y<0,
\end{eqnarray}
where the Heaviside step function $\mathbf{1}(\zeta)$ is defined as $\mathbf{1}(\zeta)= 1$  if $\zeta \ge 0$, and $\mathbf{1}(\zeta)= 0$ if $\zeta < 0$.
This means that the propagator vanishes on the negative real semi-axis when the initial value $y$ is positive, and vice versa, it vanishes on the positive real semi-axis, when the initial value $y$ is negative.
This result is derived from the fact that the function $\mathcal{D}$ is infinite when the signs of $x$ and $y$ are discordant, and then we observe that $K_0\to  0$ in this case. 
Therefore, the stochastic process cannot cross the point $x=0$, as if there were a fictitious reflecting boundary condition.  
This is consistent with the fact that both the drift and diffusion terms are zero for $x=0$. 

The short-time propagator must be now improved by means of the correction function $F$.
As before, we apply the recursive procedure stated in Eqs.(\ref{lambda}) and (\ref{dn}) and the observation of the first terms leads to the expression 
\begin{equation}
\label{dngeo}
    D_n=\frac{(-1)^nG_0^{2n}}{4^nn!}\left(2\alpha-1-\frac{H_0}{G_0^2}\right)^{2n},
\end{equation}
which must be proved by mathematical induction.
Firstly, for $n=0$, we correctly have $D_0=1$. 
Based on Eq. (\ref{dngeo}), the straightforward application of Eq. (\ref{lambda}) delivers the following coefficient $\Lambda_{n+1}$
\begin{equation}
        \Lambda_{n+1}=\frac{(-1)^{n+1}G_0^{2(n+1)}}{4^{n+1}n!}\left(2\alpha-1-\frac{H_0}{G_0^2}\right)^{2(n+1)}.
\end{equation}
For applying Eq. (\ref{dn}), we observe that $\Lambda_{n+1}$ is a constant, and we can restrict the study the case with $\sgn(x)=\sgn(y)$ for what has already been discussed in the calculation of $K_0$ (indeed, $K_0$ is zero if $\sgn(x)\neq\sgn(y)$). We then write
\begin{eqnarray}
\nonumber
        D_{n+1}&=&\frac{\Lambda_{n+1}}{\left[\sgn(x)\ln\vert\ \frac{x}{y} \vert\right]^{n+1}}\int_y^x\frac{\left[\sgn(\eta)\ln\vert\ \frac{\eta}{y} \vert\right]^{n}}{\vert\eta\vert}\d\eta\\
        \nonumber
        &=&\frac{\Lambda_{n+1}}{n+1}\\
        &=&\frac{(-1)^{n+1}G_0^{2(n+1)}}{4^{n+1}(n+1)!}\left(2\alpha-1-\frac{H_0}{G_0^2}\right)^{2(n+1)},
\end{eqnarray}
which is consistent with Eq. (\ref{dngeo}) as a result of the substitution $n\to n+1$. The mathematical induction finally proves the expression for $D_n$, which can be used to obtain the correction $F=F_0\sum_{n=0}^{+\infty}D_n(t-t_0)^n$.
We determine
\begin{equation}
    F_0=\exp\left[\int_y^x\Phi(\eta)\d\eta\right]=\exp\left[\left(-\frac{H_0}{2G_0^2}+\frac{2\alpha-3}{2}\right)\ln\vert\frac{x}{y}\vert\right],
\end{equation}
and
\begin{eqnarray}
\nonumber
    F&=&F_0\sum_{n=0}^{+\infty}\frac{(-1)^nG_0^{2n}}{4^nn!}\left(2\alpha-1-\frac{H_0}{G_0^2}\right)^{2n}(t-t_0)^n\\
    &=&F_0\exp\left[-\frac{G_0^2}{4}\left(2\alpha-1-\frac{H_0}{G_0^2}\right)^2(t-t_0)\right].
    \label{series}
\end{eqnarray}
To conclude, after straightforward calculations we obtain the complete propagator $K(x,t;y,t_0)=K_0F$ in the form
\begin{eqnarray}
\nonumber
    K=\frac{\exp\left[-\frac{\left[\ln\vert \frac{x}{y} \vert+H_0(t-t_0)-(2\alpha-1)G_0^2(t-t_0)\right]^2 }{4G_0^2(t-t_0)}\right]}{ G_0\vert x \vert \sqrt{4\pi(t-t_0)}}\mathbf{1}(x),\,\,y>0,\\\label{posy} \\
    \nonumber
    K=\frac{\exp\left[-\frac{\left[\ln\vert \frac{x}{y} \vert+H_0(t-t_0)-(2\alpha-1)G_0^2(t-t_0)\right]^2 }{4G_0^2(t-t_0)}\right]}{ G_0\vert x \vert \sqrt{4\pi(t-t_0)}}\mathbf{1}(-x),\,\,y<0,\\
    \label{negy}
\end{eqnarray}
which is in full agreement with previous investigations \cite{saaty1981,giordano2023}.
A numerical result is displayed in Fig. \ref{exgeo}, where we show the fast convergence of the process. We plot the short-time propagator $K_0$, the full propagator $K$, and some approximations $K_0Q_0$, $K_0Q_1$, and $K_0Q_2$, to show the convergence $K_0Q_m\to K $, when $m\to \infty$. Here, $Q_m=\sum_{n=0}^{m}F_n(x;y)(t-t_0)^n$, representing the partial sum of Eq. (\ref{series}).
We have rigorously proved by our method that if the geometric stochastic equation is defined on the entire real axis, then the support of the propagator consists of only the semi-axis in which the initial value is present.
The obtained result for the propagator is the log-normal distribution which is typically associated with geometric Brownian motion. We note that the result retains the dependence on the
discretization parameter $\alpha$ and thus holds for all definitions
of the the stochastic integral. However, it may be worth noting that in most applications the geometric Brownian process is considered in the It\^o-interpretation, with $\alpha = 0$.  

\subsection{A stochastic process for chromatin remodeling}
\label{secdna}

\begin{figure*}[htb]
\centering
\includegraphics[scale=0.59]{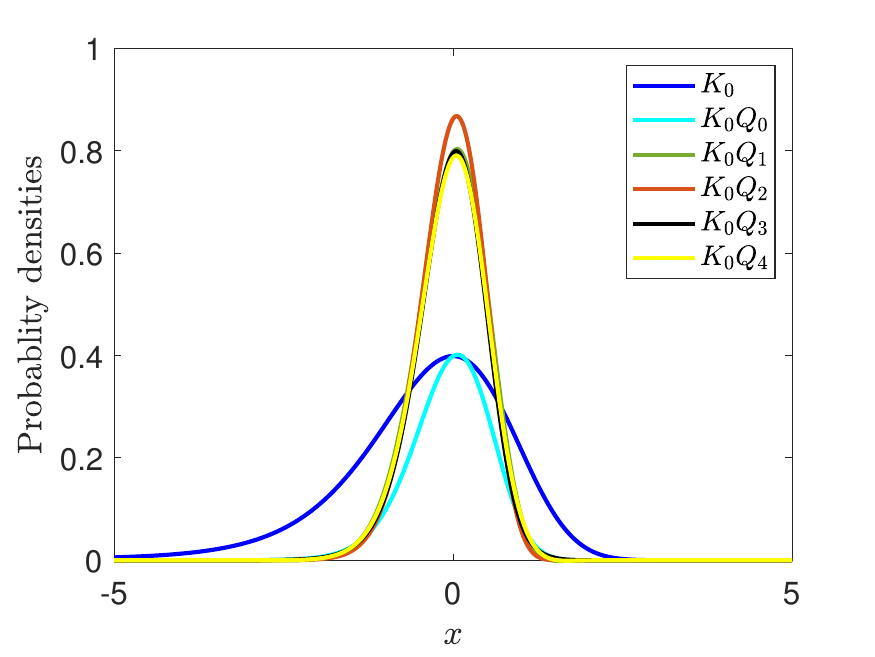}
\includegraphics[scale=0.59]{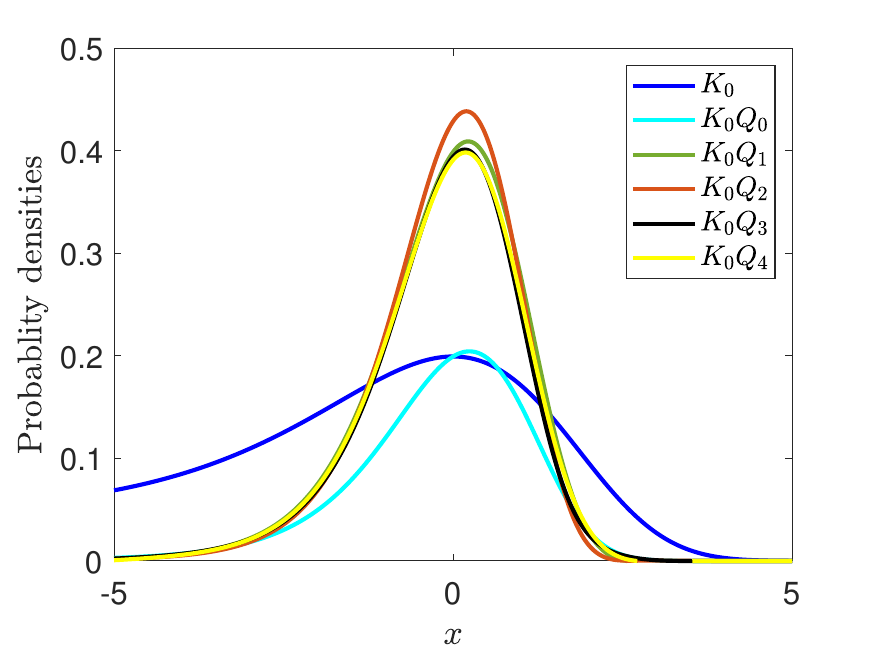}
\caption{\label{dna} Application of the recursive procedure to the stochastic equation for the remodeler-nucleosome complex. In the left panel we adopted the parameters $k=2$, $\beta=1$, $\gamma=1/10$, and $A=1$, and in the right panel $k=2$, $\beta=1$, $\gamma=1/10$, and $A=2$ (in arbitrary units). In both cases we plot the short-time propagator $K_0$, and low-order approximations $K_0Q_0$, $K_0Q_1$, $K_0Q_2$, $K_0Q_3$,and $K_0Q_4$, to show the convergence of $K_0Q_m$ for $m\to \infty$. Here, $Q_m=\sum_{n=0}^{m}F_n(x;y)(t-t_0)^n$ represents the partial sum of Eq. (\ref{serF}). In both panels we adopted $t_0=0$, and $t=1/2$. }
\end{figure*}

We will now introduce two practical applications for which the full propagator is difficult to obtain, therefore the approximation of the short-term expression of $K$ may have an important, heuristic value.

We firstly consider a stochastic model that has been used to map the stochastic motion of a chromatin remodeler acting on a nucleosome
to an active Brownian dimer \cite{kumar2008,baule2008,blossey2019}.
Originally developed for general dimeric motors, it has recently been linked via structural biology insights on the action of two 
motor domains of a chromatin remodeling enzyme on the DNA wrapped around a nucleosome. In this understanding, first one of the 
motor domain pulls in DNA into the nucleosome, and subsequently the
second motor domain expels the surplus DNA outward, thereby establishing a proper wrapping of 147 base pairs of DNA around the
protein core of the nucleosome. The motion of the two-domain nucleosome-remodeler complex can then be separated into
a center of mass motion, and a stochastic differential equation
for the relative motion, which is given by a very simple expression
in view of the complexity of the system
\begin{equation}
    \frac{\d x}{\d t}= h(x) + g(x) \xi(t)
\end{equation}
with the drift and diffusion terms given by 
\begin{eqnarray}
h(x)=-\left(\frac{1}{\beta}+\frac{1}{\beta+\gamma x}\right)k x,
\end{eqnarray}
\begin{eqnarray}
g(x)=\frac{A}{(\beta+\gamma x)^2}.
\end{eqnarray}
These expressions are valid when only active, ATP-dependent fluctuations are retained and thermal fluctuations are ignored. 
The equation is studied under the Fisk-Stratonovich interpretation ($\alpha=1/2$).
This choice was made only to fix ideas, and any other value for $\alpha$ could have been considered. In physical and biophysical applications, however, this is the value typically adopted.

The parameter $A$ signifies the strength of the active process, while the constants $\beta$ and $\gamma$ are friction terms ($\gamma\ll\beta$); $k$ is the spring constant of
a harmonic interaction potential between the remodeler motor 
domains on DNA. We stress that this model is purely phenomenological and therefore also other forms of these functions can be adopted and can in fact be found in the literature \cite{kumar2008,baule2008}.

Applying our formal analysis, we fix the initial position $y=0$, and the initial time $t_0=0$. First of all, we determine the quantity $\mathcal{D}$ as
\begin{equation}
    \mathcal{D}=\frac{1}{A}\left(\frac{1}{3}\gamma^2 x^3+\beta\gamma x^2+\beta^2 x\right),
\end{equation}
and therefore we obtain the short-time propagator in the form
\begin{equation}
     K_0(x,t;y,t_0)=\beta^2\frac{\exp\left[-\frac{1}{4A^2t}\left(\frac{1}{3}\gamma^2 x^3+\beta\gamma x^2+\beta^2 x\right)^2\right]}{  \sqrt{4\pi A^2t}}.
\end{equation}
Moreover, the first correction $F_0$ is found as
\begin{eqnarray}
\nonumber
F_0&=&\left(\frac{\gamma x+\beta}{\beta}\right)^2
\exp\left(-\frac{\beta^3 k x^2}{2A^2}-\frac{7\beta^2\gamma k x^3}{6A^2}\right.\\
&&\left.-\frac{9\beta\gamma^2 k x^4}{8A^2}-\frac{\gamma^3 k x^5}{2A^2}-\frac{\gamma^4 k x^6}{12\beta A^2}\right).
\end{eqnarray}
The application of the recursive procedure given by Eqs.(\ref{lambda}) and (\ref{dn}) yields closed form expressions for the coefficients $D_n$. While these quite lengthy expressions can be easily obtained through standard symbolic environments like {\sc MAPLE}, we display in the Appendix only the series expansion of the coefficients $D_n$ with respect to the variable $x$, keeping fixed the other physical parameters. 

The plot of the successive approximations for the propagator can be found in the left panel of Fig. \ref{dna} for a first set of parameters. Here we represented the functions  $K_0$, and some approximations $K_0Q_0$, $K_0Q_1$, $K_0Q_2$, $K_0Q_3$,and $K_0Q_4$, to show the convergence of $K_0Q_m$ for $m\to \infty$. Here, $Q_m=\sum_{n=0}^{m}F_n(x;y)(t-t_0)^n$ represents the partial sum of Eq. (\ref{serF}). It means that $Q_0=K_0$, $Q_1=F_0(1+D_1t)$, $Q_2=F_0(1+D_1t+D_2t^2)$, $Q_3=F_0(1+D_1t+D_2t^2+D_3t^3)$, and $Q_4=F_0(1+D_1t+D_2t^2+D_3t^3+D_4t^4)$. 
Here, we adopted $t_0=0$, and $t=1/2$.
It is interesting to note that a rather small number of terms is sufficient to have reasonably convergent behaviour. This can be seen from the fact that the last two approximations are almost overlapping and also from the fact that they are automatically well normalized. We emphasize that the first approximation $K_0F_0$ is again so poorly approximate that it is not even normalized.

The results for a second set of parameters, where we adopted a larger intensity of the noise related to the ATP-consumption, can be found in the right panel of Fig. \ref{dna}.
In this case, we observe a larger variance of the variable $x$, which induces larger propagator spread.
Also in this case we can remark a good convergence with a quite limited number of terms. 
This means that the recursive procedure proposed in this paper can be effectively applied to physical and biological problems of some interest with relative simplicity. In particular, it makes it possible to obtain good approximations of the propagator of one-dimensional Fokker-Planck equations even when such equations are very difficult to solve by other approaches.   

\subsection{Exponential heterogeneous diffusion}
\label{secexp}

Our final example in this Section deals with the heterogeneous exponential diffusion described by $g(x)=G_0\exp(\gamma x)$, with $G_0>0$ and $\gamma$ real. Such an exponential behavior of the diffusion coefficient has been used in a different biological context: it had been introduced to give a stochastic interpretation of the movement of the parasitic nematode {\it Phasmarhabditis hermaphrodita} \cite{hapca2007,hapca2007b,hapca2009}. 
Another biological example is given by an exponential rate of morphogen degradation in a reaction-subdiffusion model for cell development \cite{buste2010}. Morphogens are special signaling molecules whose spatial distribution controls the growth of  embryonic cells \cite{crick1970}.  
However, such biologically-motivated examples are not the only ones for which an exponential diffusion coefficient can be used. For example, they were similarly employed to model the diffusion of impurities induced by irradiation \cite{kowall1976,kesarev2009}: in this case, the exponential dependence originates from the decay of the radiation propagating in the material under study. 
Finally, they were also applied to the description of grain-boundary diffusion in nanostructured materials \cite{kesarev2015}.

We thus consider the differential equation with exponential diffusion:
\begin{equation}
    \frac{\d x}{\d t}=G_0\exp(\gamma x)\xi(t),
\end{equation}
where at first an arbitrary stochastic interpretation is assumed. 
We start our analysis by applying Eqs.(\ref{effe}), (\ref{elle}), and (\ref{phiti}), which lead to the expressions
\begin{eqnarray}
f(x)&=&4(\alpha-1)G_0^2\gamma^2\exp(2\gamma x),
\\ \ell(x)&=&2(\alpha-2)G_0^2\gamma\exp(2\gamma x),\\
\Phi(x)&=&\frac{\gamma}{2}(2\alpha-3). 
\end{eqnarray}
As done previously, we applied the recursion for the first terms manually, see Eqs.(\ref{lambda}) and (\ref{dn}), and realized that the parameters $D_n$ are described by the relation
\begin{equation}
    D_n=\frac{(-1)^nG_0^{2n}\gamma^{2n}\exp\left[n\gamma(x+y)\right]}{4^n n!}\prod_{k=0}^{n-1}\left[4\alpha^2-(2k+1)^2\right],
    \label{dnprod}
\end{equation}
which we now demonstrate by mathematical induction.
Firstly, for $n=0$ we have $D_0=1$, considering that the empty product is unitary.
Next, let's assume that the formula is true
for a natural number $n$, and since ${\d D_n}/{\d x}=n\gamma D_n$, and $\d \Phi/\d x=0$, we get from Eq. (\ref{lambda})
\begin{equation}
    \Lambda_{n+1}=-G_0^2\gamma^2\exp(2\gamma x)D_n\frac{4\alpha^2-(2n+1)^2}{4}.
\end{equation}
We introduce the polynomials $Q_n(\zeta)$ such that $Q_0(\zeta)=1$ and $Q_{n+1}(\zeta)=[4\zeta^2-(2n+1)^2]Q_n(\zeta)$.
They allow us to write Eq. (\ref{dnprod}) in the form
\begin{equation}
    D_n=\frac{(-1)^nG_0^{2n}\gamma^{2n}\exp\left[n\gamma(x+y)\right]}{4^n n!}Q_n(\alpha),
    \label{dnpoli}
\end{equation}
Now, by means of the expression  $\mathcal{D}=[\exp(-\gamma y)-\exp(-\gamma x)]/{G_0\gamma}$, we can apply Eq. (\ref{dn}) and we obtain after straightforward calculation
\begin{equation}
D_{n+1}=\frac{(-1)^{n+1}G_0^{2(n+1)}\gamma^{2(n+1)}\exp\left[(n+1)\gamma(x+y)\right]}{4^{n+1}(n+1)!}Q_{n+1}(\alpha), 
\end{equation} 
which finally proves Eq. (\ref{dnprod}) by mathematical induction.

In this case we have obtained the closed form of the Taylor coefficients but it is important now to remark that the corresponding series leads to a meaningful result only if  $\alpha=1/2$. In fact, with $\alpha=1/2$, all coefficients $D_n$ {\it vanish} with $n\ge 1$ as seen by observing the product in Eq. (\ref{dnprod}). The expansion is thus cut off after the first term. 
On the other hand, it is seen that with $\alpha\neq 1/2$ the series of powers with coefficients $D_n$ cannot converge because of the divergent behavior of $D_n$ with $n$ since the product grows  roughly as $(n!)^2$.
Therefore, only the Fisk-Stratonovich interpretation leads to a 
well-defined development for the exponential diffusion (this case will be further discussed in the next Section).
As a consequence the probability density for the exponential diffusive process cannot be developed in a Taylor series independently from the interpretation of the stochastic integrals. 

Although it is interesting to have understood that the solution by series can be obtained only for certain values of $\alpha$, a theoretical understanding of this fact is lacking. In particular, for the time being it is impossible to predict for a certain problem what are the values of $\alpha$ for which we have a convergent solution. This point certainly requires further theoretical analysis in the near future. 

\section{Stochastic processes in the Fisk-Stratonovich interpretation - new insights from our formalism}
\label{secfin}

In this Section we take a different point of view. Rather than 
starting out from a specifically defined model, as we did in Section III, we here develop 
rather general consequences following from of our formalism. In this Section
we consider stochastic processes always in the Fisk-Stratonovich calculus. We first report on some general results in the case of pure diffusions, i.e. in the absence of a drift term. Then we turn to the special case in which drift and diffusion terms conspire in such a way as to render our recursion equations particularly simple to solve. 

\subsection{The case of pure diffusion and no drift}

We first consider a stochastic process without drift, $h(x)=0$, interpreted through the Fisk-Stratonovich integration with $\alpha=1/2$. We have
\begin{equation}
\label{lgpart}
    \frac{\d x}{\d t}=g(x)\xi(t),
\end{equation}
with $g(x)\ge 0$ for any $x\in\mathbb{R}$.
To take advantage of our procedure, we can directly apply Eqs.(\ref{effe}), (\ref{elle}), and (\ref{phiti}) and obtain the expressions
\begin{eqnarray}
f(x)&=&-g'^2(x)-g(x)g''(x),\\ \ell(x)&=&-3g(x)g'(x),\\
\label{phipart}
\Phi(x)&=&-\disp\frac{g'(x)}{g(x)}. 
\end{eqnarray}
In order to implement the recursion, we start form $D_0=1$ and we calculate $\Lambda_1$ through Eq. (\ref{lambda}), eventually obtaining
\begin{eqnarray}
\nonumber
        \Lambda_1(x)&=&g'^2(x)+g(x)g''(x)-3g(x)g'(x)\frac{g'(x)}{g(x)}\\
        \nonumber
        &&+g^2(x)\left(-\frac{g''(x)g(x)-g'^2(x)}{g^2(x)}+\frac{g'^2(x)}{g^2(x)}\right)\\
        \nonumber
        &=&g'^2(x)+g(x)g''(x)-3g'^2(x)-g''(x)g(x)\\
        &&+g'^2(x)+g'^2(x)=0.
 \end{eqnarray}
The fact that $\Lambda_1=0$, by induction, implies that all parameters $\Lambda_n$  and $D_n$  are zero from $n=1$ until infinity. Hence, the correction function $F$ is simply given by the first term $F_0$, which can be calculated as follows
\begin{equation}
    F_0(x;y)=\exp\left[\int_y^x\Phi(\eta,y)\d\eta \right]=\frac{g(y)}{g(x)},
\end{equation}
where we used Eq. (\ref{phipart}).
To conclude, we obtain the closed form expression of the propagator for Eq. (\ref{lgpart}) as $K=K_0F_0$, which turns out to be
\begin{equation}
    K(x,t;y,t_0)=\frac{\exp\left[-\frac{1}{4(t-t_0)}\left(\int_y^x\frac{\d\eta}{g(\eta)}\right)^2\right]}{g(x)\sqrt{4\pi(t-t_0)}}.
    \label{propar}
\end{equation}
The Wiener process is of course retrieved when $g(x)$ is a constant.

The fact that in this specific case we were able to find a rather general solution should not be too surprising.  
In fact, the stochastic differential equation is interpreted by Fisk-Stratonovich integration and thus all the rules of mathematical analysis remain unchanged and applicable as if the equation were not stochastic. 
Based on this observation, we see that the simple transformation law $z(x)=\int_a^x d\eta/g(\eta)$ (for some real constant $a$) is able to convert the starting equation into $\d z/\d t=\xi(t)$, which describes a simple Wiener process. This alternative approach leads indeed to the same result stated in Eq. (\ref{propar}). 
An interesting investigation on this type of equation can be found in Ref. \cite{fa2005}. 

\begin{figure}[t]
\centering
\includegraphics[scale=0.60]{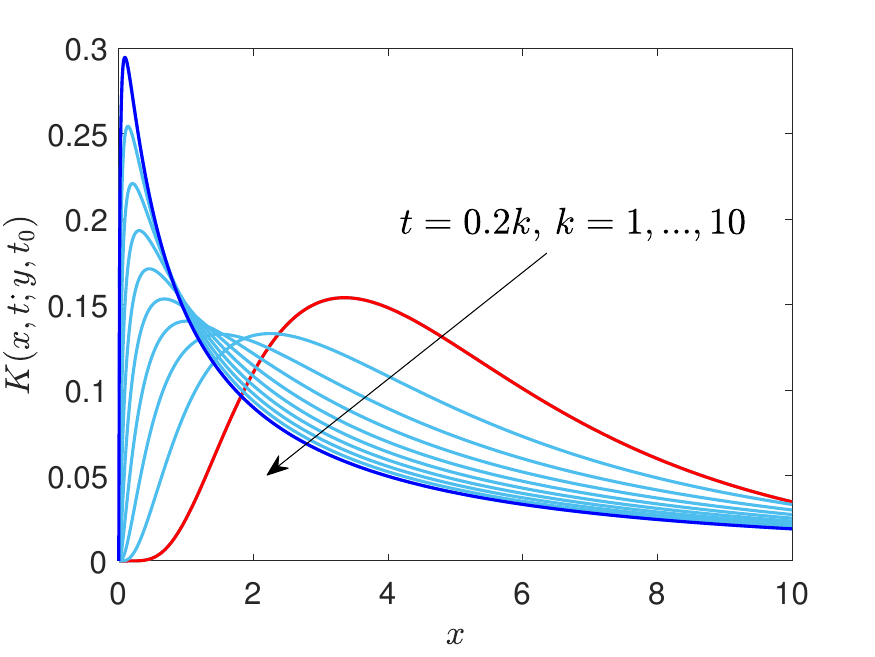}
\caption{\label{exgeo2} Example of time evolution for a geometric Brownian process. We plot the full propagator $K$ given in Eq. (\ref{expfig}) for different values of the time $t=0.2 k$, with $k=1,...,10$ (start time in red and end time in blue). We adopted the parameters, $G_0=1$, $y=5$, $t_0=0$, and $\alpha=1/2$ (in arbitrary units).}
\end{figure}

\begin{figure*}[t]
\centering
\includegraphics[scale=0.60]{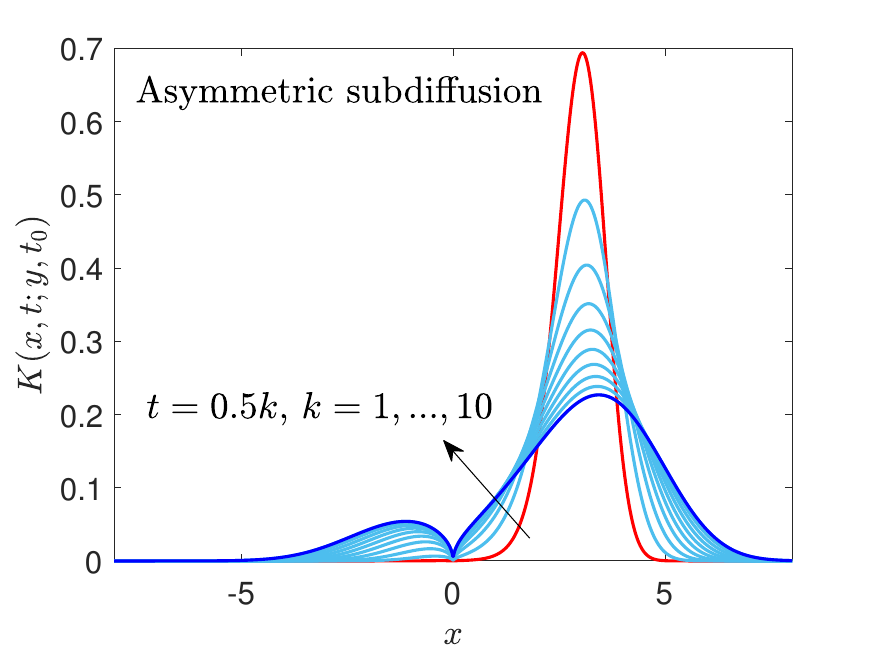}
\includegraphics[scale=0.60]{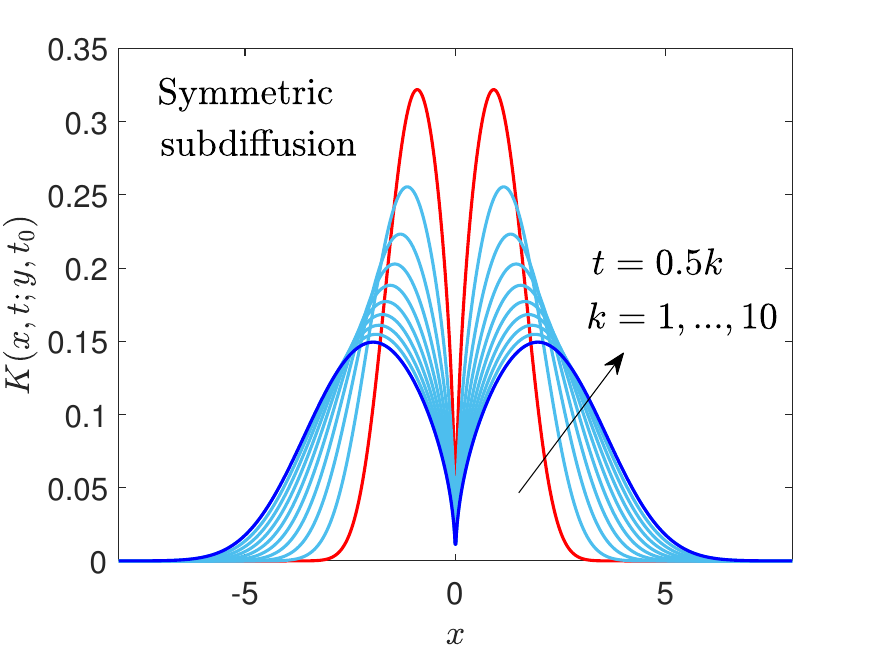}
\caption{\label{sub} Asymmetric and symmetric subdiffusion processes. We plot the full propagators $K$ given in Eqs.(\ref{asymdif}) and (\ref{denspec}) for different values of the time $t=0.5 k$, with $k=1,...,10$ (start time in red and end time in blue), and for $\beta=-1/2$ ($e=2/3$). We adopted the parameters, $G_0=1$, $y=3$ (for the asymmetric subdiffusion, left panel), $y=0$ (for the symmetric subdiffusion, right panel), $t_0=0$, and $\alpha=1/2$ (in arbitrary units).}
\end{figure*}

\begin{figure*}[t]
\centering
\includegraphics[scale=0.60]{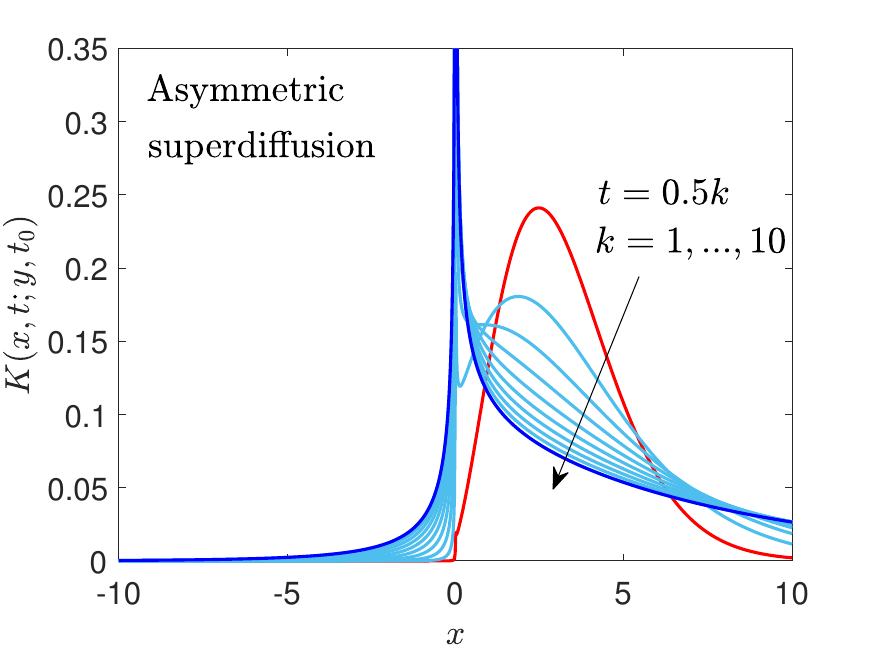}
\includegraphics[scale=0.60]{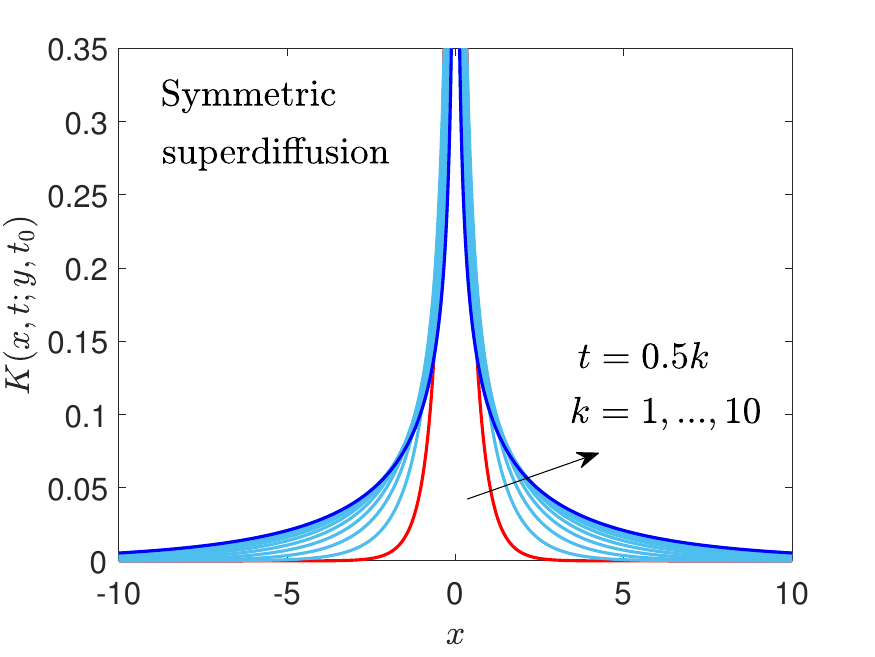}
\caption{\label{sup} Asymmetric and symmetric superdiffusion processes. We plot the full propagators $K$ given in Eqs.(\ref{asymdif}) and (\ref{denspec}) for different values of the time $t=0.5 k$ (start time in red and end time in blue), with $k=1,...,10$, and for $\beta=+1/2$ ($e=2$). We adopted the parameters, $G_0=1$, $y=3$ (for the asymmetric superdiffusion, left panel), $y=0$ (for the symmetric superdiffusion, right panel), $t_0=0$, and $\alpha=1/2$ (in arbitrary units).}
\end{figure*}

This problem is useful to further explore the behavior of several systems with heterogeneous diffusion. For example, if we consider the geometric Brownian motion case from Section III with $g(x)=G_0\vert x\vert$ but without drift, we simply obtain
\begin{eqnarray}
\label{expfig}
    K(x,t;y,t_0)&=&\frac{\exp\left[-\frac{\ln^2\vert \frac{x}{y} \vert }{4G_0^2(t-t_0)}\right]}{ G_0\vert x \vert \sqrt{4\pi(t-t_0)}}\mathbf{1}(x),\,\,\,\,\,\,y>0,\\ 
    K(x,t;y,t_0)&=&\frac{\exp\left[-\frac{\ln^2\vert \frac{x}{y} \vert }{4G_0^2(t-t_0)}\right]}{ G_0\vert x \vert \sqrt{4\pi(t-t_0)}}\mathbf{1}(-x),\,\,\,\,\,\,y<0,
\end{eqnarray}
in full agreement with Eqs.(\ref{posy}) and (\ref{negy}), with $H_0=0$ and $\alpha=1/2$. 
The geometric Brownian motion without drift, in the case of Fisk-Stratonovich interpretation, can thus be studied directly with this simple approach, without developing the full procedure as in the previous Section \ref{appgeo}. An example can be found in Fig. \ref{exgeo2}, where we plot the time evolution of the propagator given in Eq. (\ref{expfig}), with a positive initial condition. 

A further interesting case concerns the diffusive behavior described by the power law $g(x)=G_0\vert x\vert^\beta$, where we assume that $\beta<1$ for reasons that will be clarified shortly. We have to study the integral $\int_y^x \d\eta/\vert \eta \vert^\beta$. 
It is rather simple to verify the validity of the indefinite integral $\int \d\eta/\vert \eta \vert^\beta=\sgn(\eta)\vert\eta\vert^{1-\beta}/(1-\beta)+C$ for some real constant $C$.
Hence, the hypothesis $\beta<1$ allows us to state that the improper integral is convergent in the neighborhood of $\eta=0$ (indeed, $1-\beta>0$), and we can write
\begin{eqnarray}
\nonumber
    \int_y^x \frac{\d\eta}{\vert \eta \vert^\beta}&=&\sgn(x)\frac{\vert x\vert^{1-\beta}}{1-\beta}-\sgn(y)\frac{\vert y\vert^{1-\beta}}{1-\beta}\\
    &=&\frac{\vert x\vert^{2-\beta}}{x(1-\beta)}-\frac{\vert y\vert^{2-\beta}}{y(1-\beta)},
\end{eqnarray}
which is valid for any real numbers $x$ and $y$.
Unlike the case of geometric Brownian motion, we find here a propagator defined on the entire real axis (with $\beta<1$)
\begin{equation}
    K(x,t;y,t_0)=\frac{\exp\left[-\frac{1}{4G_0^2(t-t_0)(1-\beta)^2}\left(\frac{\vert x\vert^{2-\beta}}{x}-\frac{\vert y\vert^{2-\beta}}{y}\right)^2\right]}{ G_0\vert x \vert^\beta \sqrt{4\pi(t-t_0)}}.
    \label{asymdif}
\end{equation}
This result represents a generalization of the expression obtained in Ref. \cite{Cherstvy2013} for $y=0$, which reads as
\begin{equation}
    K(x,t;y,t_0)=\frac{\exp\left[-\frac{1}{4G_0^2(t-t_0)(1-\beta)^2}\vert x\vert^{2(1-\beta)}\right]}{ G_0\vert x \vert^\beta \sqrt{4\pi(t-t_0)}}.
    \label{denspec}
\end{equation}
This process is called subdiffusive if $\beta<0$ and superdiffusive if $0<\beta<1$. 
This terminology derives from the calculation of mean-squared displacement from Eq. (\ref{denspec}), resulting in $\mathbb{E}(x^2(t))\sim t^{1/(1-\beta)}$, where we identify the exponent $e=1/(1-\beta)$. 
Depending on the value of the
anomalous diffusion exponent $e$ we can thus distinguish subdiffusion
($0<e<1$), superdiffusion ($e>1$), standard Brownian motion ($e=1$),  ballistic wave-like motion ($e=2$) and geometric Brownian motion ($e\to\infty$).
This analysis shows that the anomalous diffusion observed  in several systems can be explained by means of a heterogeneous diffusion, for example of the power-law type. Examples of asymmetric ($y\neq 0$) and symmetric ($y=0$) subdiffusion and superdiffusion processes can be found in Figs.\ref{sub} and \ref{sup}, respectively. In these cases the mean-squared displacement exponent assumes the values $e=2/3$ (subdiffusive behavior) and $e=2$ (superdiffusive behavior).

\begin{figure}[t]
\centering
\includegraphics[scale=0.60]{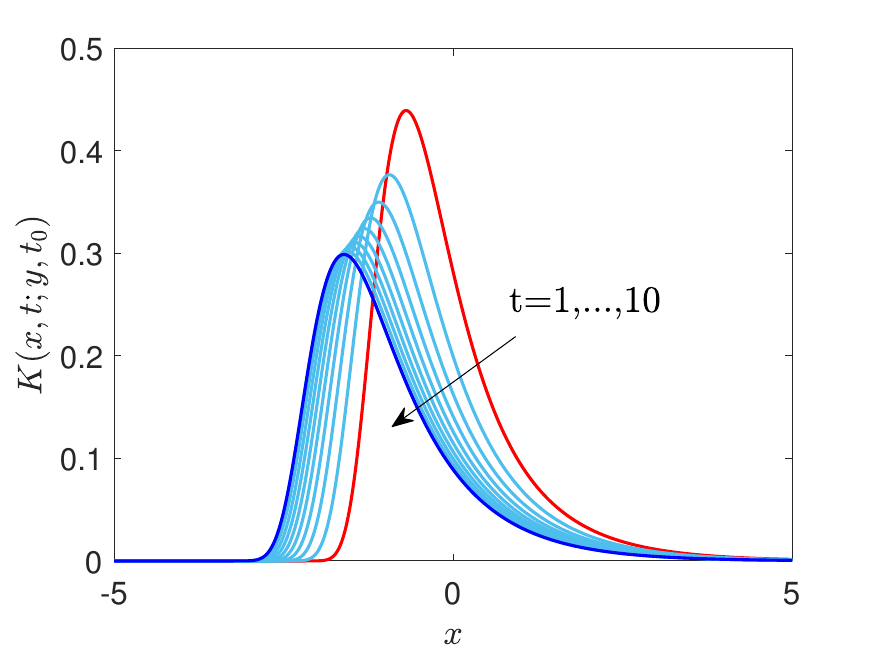}
\caption{\label{exexp} Example of time evolution for a exponential process. We plot the full propagator $K$ given in Eq. (\ref{proparexp}) for different values of the time $t=1,...,10$ (start time in red and end time in blue). We adopted the parameter values $G_0=1$, $\gamma=1$, $y=0$, $t_0=0$, and $\alpha=1/2$ (in arbitrary units).}
\end{figure}

This procedure also works, of course, for the case of heterogeneous
exponential diffusion, already discussed in Section \ref{secexp}. 
The application of Eq. (\ref{propar}) leads directly to the propagator in the form
\begin{equation}
    K(x,t;y,t_0)=\frac{\exp\left[-\frac{1}{4G_0^2\gamma^2(t-t_0)}\left(\exp(-\gamma y)-\exp(-\gamma x)\right)^2\right]}{G_0\exp(\gamma x)\sqrt{4\pi(t-t_0)}},
    \label{proparexp}
\end{equation}
in agreement with the result in Ref. \cite{Cherstvy2013}.
In this case, the mean-squared displacement shows a logarithmic time dependence, which is a characteristic feature of ultraslow processes \cite{Cherstvy2013}.
The existence of the solution for the case with the exponential heterogeneous diffusion under the Fisk-Stratonovich interpretation is consistent with the result of Section \ref{secexp}, affirming that the development in a power series is possible 
only if $\alpha=1/2$. 
An example can be seen in Fig. \ref{exexp}, where we show the time behavior of the propagator for the exponential process with $G_0=1$ and $\gamma=1$. 
We observe that the probability density evolves slowly in the direction where the diffusion coefficient is lower, being driven
by the gradient of fluctuations. 

Although the analysis developed here concerns the Fisk-Stratonovich interpretation, some generalizations can be mentioned. Indeed, it is interesting to remark that Eq.(\ref{lgpart}), interpreted with $\alpha=1/2$, is equivalent to the equation
\begin{equation}
\label{lgparta}
    \frac{\d x}{\d t}=(1-2\alpha)g(x)\frac{\d g(x)}{\d x}+ g(x)\xi(t),
\end{equation}
interpreted with an arbitrary value of $\alpha$.
Then the solution given in Eq. (\ref{propar}) is valid also in these cases, that are outside the Fisk-Stratonovich interpretation.

\subsection{Stochastic processes with drift under the Fisk-Stratonovich interpretation}

We now return to stochastic processes with drift, studied under the Fisk-Stratonovich interpretation, and search for functions $h$ and $g$ that are able to significantly simplify the recursive procedure obtained with our approach. In particular, we look for the form of these functions that generate constant coefficients $\Lambda_n$ and $D_n$ (i.e., independent of $x$). 
From Eq.(\ref{lambda}), we see this can be true only if 
\begin{eqnarray}
-f-\ell\Phi+g^2\frac{\partial \Phi}{\partial x}+g^2\Phi^2=C,
\label{condgen}
\end{eqnarray}
where $C$ is a constant.
By using Eqs.(\ref{effe}), (\ref{elle}), and (\ref{phiti}) for $f$, $\ell$,  and $\Phi$, this condition can be rewritten as
\begin{eqnarray}
\nonumber
&&h'+2(\alpha-1)h\frac{g'}{g}+\frac{h^2}{2g^2}\\
&&\,\,\,\,\,\,\,\,\,\,\,\,=-2C-(2\alpha-1)\left[\frac{2\alpha-1}{2}(g')^2+gg''\right],
\label{condn}
\end{eqnarray}
where $\alpha$ is the stochastic discretization parameter.
This relation must be fulfilled by $h$ and $g$ in order to have constant coefficients $\Lambda_n$ and $D_n$. 
In the particular case with $h=0$ and $\alpha=1/2$, we get $C=0$, and therefore we retrieve the results of the previous Section (all coefficients are zero for $n\ge 1$).
We now want to retain the Fisk-Stratonovich interpretation but also assume that $h\neq 0$.
In this situation, Eq.(\ref{condn}) reduces to
\begin{eqnarray}
&&h'-h\frac{g'}{g}+\frac{h^2}{2g^2}=-2C,
\label{eqfisk}
\end{eqnarray}
A simple assumption that allows us to verify this condition is given by $h(x)=-mg(x)$, where $m$ is a real parameter.
It means that we are studying the stochastic differential equation
\begin{equation}
    \frac{\d x}{\d t}=-mg(x)+g(x)\xi(t),
    \label{eqhg}
\end{equation}
where the drift term is proportional to the diffusion term. In this particular case we get
\begin{eqnarray}
f(x)&=&-mg'-(g')^2-gg'',\\
\ell(x)&=&-mg-3gg',\\
\Phi(x)&=&-\frac{1}{2}\frac{m}{g}-\frac{g'}{g}.
\end{eqnarray}
The short-time propagator is given by the expression
\begin{eqnarray}
    K_0(x,t;y,t_0)&=&\frac{\exp\left[-\frac{1}{4(t-t_0)}\left(\int_y^x\frac{\d\eta}{ g(\eta) }\right)^2\right]}{ g(y) \sqrt{4\pi(t-t_0)}}, .
\end{eqnarray}
The first correction term $F_0$ can be calculated as follows
\begin{equation}
    F_0=\exp\left[\int_y^x\Phi(\eta)\d\eta \right]=\frac{g(y)}{g(x)}\exp\left[-\frac{m}{2}\int_y^x\frac{\d\eta}{ g(\eta) }\right],
\end{equation}
and the other coefficients must be determined through the iteration stated in Eqs.(\ref{lambda}) and (\ref{dn}).
In this case, Eq.(\ref{condgen}) is fulfilled with a constant $C=-m^2/4$, and therefore we have from Eq.(\ref{lambda}) 
\begin{equation}
    \Lambda_n=-\frac{1}{4}m^2D_{n-1}.
\end{equation}
When this relation is substituted into Eq.(\ref{dn}), the following simple recursion is found
\begin{equation}
    D_n=-\frac{1}{4n}m^2D_{n-1},
\end{equation}
which leads to the result
\begin{equation}
    D_n=(-1)^n\frac{m^{2n}}{4^n n!}
\end{equation}
that complies with the condition $D_0=1$.
The correction function $F$ can be summed as follows
\begin{equation}
F=F_0\sum_{n=0}^{+\infty}\frac{1}{n!}\left[-\frac{m^2(t-t_0)}{4}\right]^n=F_0\exp\left[-\frac{m^2}{4}(t-t_0)\right],
\end{equation}
and the propagator $K=K_0F$ is finally found in the form
\begin{equation}
    K(x,t;y,t_0)=\frac{\exp\left[-\frac{1}{4(t-t_0)}\left(\int_y^x\frac{\d\eta}{g(\eta)}+m(t-t_0)\right)^2\right]}{g(x)\sqrt{4\pi(t-t_0)}}.
    \label{propargen}
\end{equation}
This result, as expected, reduce to Eq.(\ref{propar}) when $m=0$, i.e., when we set the drift to zero. 
The problem described by Eq.(\ref{eqhg}), just solved for $\alpha=1/2$, is useful for studying various stochastic population growth models. For instance, as the diffusion function $g$ varies, we can recover the Verhulst logistic model, the Gompertz model, the Shoener model, the Richards model, and the Smith model \cite{kimura1964,montroll1971,birch1999,sakanoue2007,fa2018}.

Another interesting example can be found by considering $g(x)$ fixed and by searching for the solution of Eq.(\ref{eqfisk}) with respect to $h(x)$. This equation is of Riccati type, which means that it is composed of a linear combination of the derivative of $h$, a linear term in $h$, a term independent of $h$ and finally a quadratic term in $h$. We learned from the previous example that $h(x)=-mg(x) $ is a particular solution of this equation (with $C=-m^2/4$). 
We can therefore search for a general solution of Eq.(\ref{eqfisk}) in the form $h(x)=-mg(x)+\tau(x)$ (again with $C=-m^2/4$).
A simple calculation proves that $\tau(x)$ satisfies the following equation
\begin{equation}
    \tau'=\left(\frac{g'}{g}+\frac{m}{g}\right)\tau-\frac{\tau^2}{2g^2},
    \label{bernoulli}
\end{equation}
which is of the Bernoulli type (with a structure similar to the Riccati equation without the term independent of the unknown $\tau$).
It means that the knowledge of a particular solution of Eq.(\ref{eqfisk}), in the form $h(x)=-mg(x) $, allowed us to move from a Riccati equation to a Bernoulli equation, thus slightly simpler.
A further change of variable, namely $\theta=1/\tau$, finally reduces the equation to the linear case.
Straightforward calculations lead indeed to the equation
\begin{equation}
    \theta'=-\left(\frac{g'}{g}+\frac{m}{g}\right)\theta+\frac{1}{2g^2},
    \label{linear}
\end{equation}
which can be solved without additional effort. 
In fact, by defining a function $p(x)$ such that $p'(x)=1/g(x)$, i.e. $p$ is any indefinite integral of $1/g$, we have the solution
\begin{equation}
\theta(x)=\frac{S}{g(x)}e^{-mp(x)}+\frac{1}{2g(x)}e^{-mp(x)}\int^x \frac{1}{g(\eta)}e^{mp(\eta)}d\eta,
\label{thetasol}
\end{equation}
where $S$ is an arbitrary integration constant. 
We can conclude that the general solution of Eq.(\ref{eqfisk}) is given by $h(x)=-mg(x)+1/\theta(x)$. 
As a result of the construction developed, the Fokker-Planck equation generated by $h$ and $g$ definitely has constant coefficients $D_n$ and thus its propagator can be found without difficulty. 
This defines a new class of Langevin equations for which the exact solution can be calculated.

\begin{figure*}[t]
\centering
\includegraphics[scale=0.59]{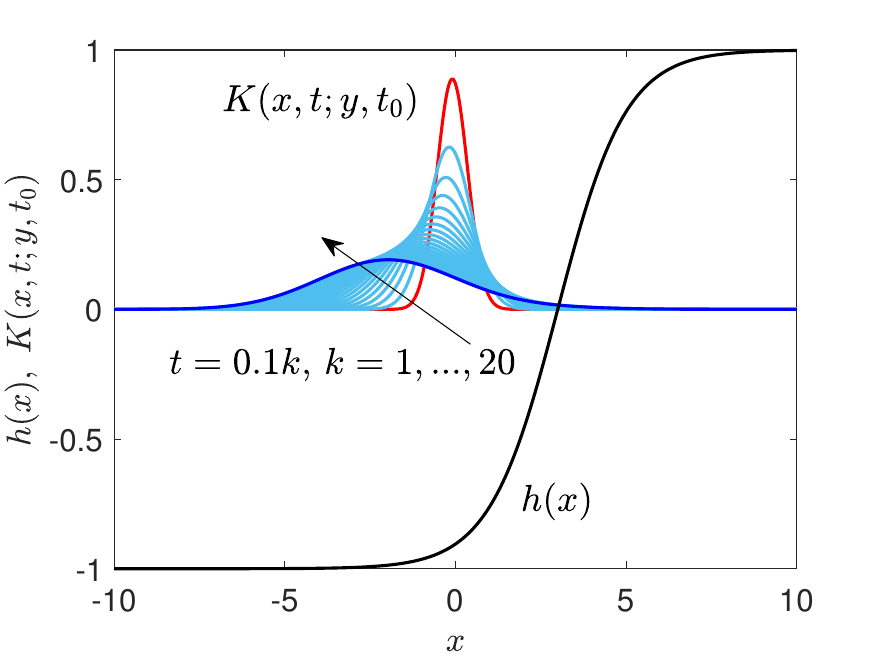}
\includegraphics[scale=0.59]{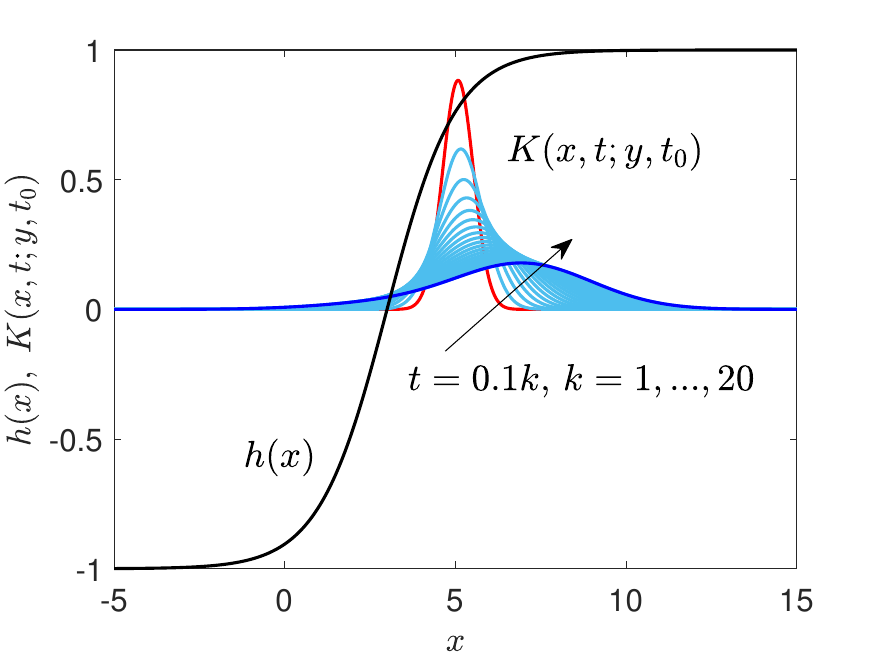}
\caption{\label{sig} Stochastic process with a sigmoidal drift. We plot the full propagators $K$ given in Eq.(\ref{sigmoidal}) for different values of the time $t=0.1 k$, with $k=1,...,20$ (start time in red and end time in blue). We also plot the sigmoidal drift $h(x)$ (black lines). We adopted the parameter values, $G_0=1$, $H_0=1$, $t_0=0$, $S=10$, and $\alpha=1/2$ (in arbitrary units).
We observe a regressive diffusion in left panel with $y=0$, and a progressive diffusion in the right panel with $y=5$. }
\end{figure*}

Let's study a specific example in detail concerning a constant diffusion term $g=G_0$. This choice leads to a drift term $h(x)=-mG_0+1/\theta(x)=-H_0+1/\theta(x)$, where $\theta(x)$ can be obtained through Eqs.(\ref{linear}) and (\ref{thetasol}). More explicitly
\begin{equation}
    h(x)=H_0\frac{1-2H_0S\exp\left(-\frac{H_0}{G_0^2}x\right)}{1+2H_0S\exp\left(-\frac{H_0}{G_0^2}x\right)},
\end{equation}
where $S $ is arbitrary. We are therefore studying the nonlinear Langevin equation with a sigmoidal drift
\begin{equation}
    \frac{\d x}{\d t}=H_0\frac{1-2H_0S\exp\left(-\frac{H_0}{G_0^2}x\right)}{1+2H_0S\exp\left(-\frac{H_0}{G_0^2}x\right)}+G_0\xi(t),
        \label{langste}
\end{equation}
interpreted through $\alpha=1/2$. For this equation, the function $\Phi(x)$ can be written as
\begin{equation}
    \Phi(x)=\frac{h(x)}{2G_0^2}=\frac{H_0}{2G_0^2}\frac{1-2H_0S\exp\left(-\frac{H_0}{G_0^2}x\right)}{1+2H_0S\exp\left(-\frac{H_0}{G_0^2}x\right)},
\end{equation}
and therefore
\begin{equation}
    F_0=\exp\left[\int_y^x\Phi(\eta)\d\eta \right]=\frac{\exp\left(\frac{H_0}{2G_0^2}x\right)+2H_0S\exp\left(-\frac{H_0}{2G_0^2}x\right)}{\exp\left(\frac{H_0}{2G_0^2}y\right)+2H_0S\exp\left(-\frac{H_0}{2G_0^2}y\right)}.
\end{equation}
\\
Moreover, the constant $C$ is given by $C=-m^2/4=-H_0^2/(4G_0^2)$.
Hence, we have
\begin{equation}
    D_n=(-1)^n\frac{m^{2n}}{4^n n!}=(-1)^n\frac{1}{4^n n!}\left(\frac{H_0}{G_0}\right)^{2n}.
\end{equation}
We can calculate the correction function $F$
\begin{equation}
F=F_0\exp\left[-\frac{1}{4}\left(\frac{H_0}{G_0}\right)^2(t-t_0)\right],
\end{equation}
and we finally have the propagator $K=K_0F$ in explicit form
\begin{eqnarray}
\nonumber
    K(x,t;y,t_0)&=&\frac{1}{\sqrt{4\pi G_0^2(t-t_0)}}\exp\left[-\frac{(x-y)^2}{4G_0^2(t-t_0)}\right]\\
    \nonumber
    &&\times\frac{\exp\left(\frac{H_0}{2G_0^2}x\right)+2H_0S\exp\left(-\frac{H_0}{2G_0^2}x\right)}{\exp\left(\frac{H_0}{2G_0^2}y\right)+2H_0S\exp\left(-\frac{H_0}{2G_0^2}y\right)}\\
    &&\times\exp\left[-\frac{1}{4}\left(\frac{H_0}{G_0}\right)^2(t-t_0)\right].
    \label{sigmoidal}
\end{eqnarray}
It can be verified that this expression is indeed the solution of the Fokker-Planck equation associated to Eq.(\ref{langste}).
A numerical example can be found in Fig. \ref{sig}, where we show the sigmoidal drift with the propagator evolution for two different initial conditions, corresponding to the positive and negative regions of the sigmoidal function. We observe a regressive diffusion in the first case and a progressive diffusion ion the second one, as expected. 
We also observe that the rate of displacement of the mean value of the variable $x$ converges to a well-determined value for long times because of the horizontal asymptotes of the sigmoidal function. 
Moreover, the variance is an increasing function due to the noise fluctuations. 

Another example deals with the heterogeneous exponential diffusion of the form $g(x)=G_0\exp(\gamma x)$. 
We consider again $\alpha=1/2$, and we search for the function $h$ given by $h(x)=-mg(x)+1/\theta(x)$, as before. We define $H_0=mG_0$, and we obtain $p(x)=-\exp(-\gamma x)/(G_0 \gamma)$, such that $p'(x)=1/g(x)$.
The application of previous procedure delivers
\begin{eqnarray}
\nonumber
    h(x)&=&H_0\exp(\gamma x)\frac{1-2\frac{H_0}{G_0}S\exp\left[-\frac{H_0}{G_0}p(x)\right]}{1+2\frac{H_0}{G_0}S\exp\left[-\frac{H_0}{G_0}p(x)\right]}\\
    &=&H_0\exp(\gamma x)\frac{1-2\frac{H_0}{G_0}S\exp\left[\frac{H_0}{G_0^2\gamma}\exp(-\gamma x)\right]}{1+2\frac{H_0}{G_0}S\exp\left[\frac{H_0}{G_0^2\gamma}\exp(-\gamma x)\right]},\,\,\,\,\,\,\,\,
\end{eqnarray}
where $S$ is an arbitrary constant. 
It means that we study the nonlinear Langevin equation with heterogeneous diffusion
\begin{eqnarray}
\nonumber
    \frac{\d x}{\d t}&=&H_0\exp(\gamma x)\frac{1-2\frac{H_0}{G_0}S\exp\left[\frac{H_0}{G_0^2\gamma}\exp(-\gamma x)\right]}{1+2\frac{H_0}{G_0}S\exp\left[\frac{H_0}{G_0^2\gamma}\exp(-\gamma x)\right]}\\
    &&+G_0\exp(\gamma x)\xi(t),
\end{eqnarray}
under the Fisk-Stratonovich interpretation. 
With straightforward calculation we find
\begin{eqnarray}
F_0&=&\frac{1+2\frac{H_0}{G_0}S\exp\left[\frac{H_0}{G_0^2\gamma}\exp(-\gamma x)\right]}{1+2\frac{H_0}{G_0}S\exp\left[\frac{H_0}{G_0^2\gamma}\exp(-\gamma y)\right]}\\
\nonumber
&&\times \exp\left\lbrace-\gamma(x-y)-\frac{H_0}{2G_0^2\gamma}\left[\exp(-\gamma x)-\exp(-\gamma y)\right]\right\rbrace,
\end{eqnarray}
and the correction function $F$ appears to be
\begin{equation}
F=F_0\exp\left[-\frac{1}{4}\left(\frac{H_0}{G_0}\right)^2(t-t_0)\right].
\end{equation}
To conclude, we can obtain the short time propagator as
\begin{eqnarray}
    K_0=\frac{\exp\left\lbrace-\frac{\left[\exp(-\gamma x)-\exp(-\gamma y)\right]^2}{4G_0^2\gamma ^2(t-t_0)}\right\rbrace}{\sqrt{4\pi G_0^2\exp(2\gamma y)(t-t_0)}},
\end{eqnarray}
and the complete propagator in the form
\begin{eqnarray}
\nonumber
    K&=&\frac{\exp\left\lbrace-\frac{\left[\exp(-\gamma x)-\exp(-\gamma y)\right]^2}{4G_0^2\gamma ^2(t-t_0)}\right\rbrace}{\sqrt{4\pi G_0^2\exp(2\gamma y)(t-t_0)}}\exp\left[-\frac{1}{4}\left(\frac{H_0}{G_0}\right)^2(t-t_0)\right]\\
    &&\times\frac{1+2\frac{H_0}{G_0}S\exp\left[\frac{H_0}{G_0^2\gamma}\exp(-\gamma x)\right]}{1+2\frac{H_0}{G_0}S\exp\left[\frac{H_0}{G_0^2\gamma}\exp(-\gamma y)\right]}\\
\nonumber
&&\times \exp\left\lbrace-\gamma(x-y)-\frac{H_0}{2G_0^2\gamma}\left[\exp(-\gamma x)-\exp(-\gamma y)\right]\right\rbrace.
\end{eqnarray}
This is the final result concerning the exponential diffusion with a specific drift and it can be checked by direct substitution into the Fokker-Planck equation.

This procedure, once fixed an arbitrary heterogeneous diffusion described by $g(x)$, enables us to find a function $h(x)$ that allows solving the associated Fokker-Planck equation with some degree of ease, as shown in the examples above. 
This can be useful for constructing stochastic systems whose solutions are known, with a variety of practical applications.

\section{Conclusions}
\label{secconc}

In this paper we have derived a short-time expansion for the
probability distribution of a Fokker-Planck equation. We have
shown that its kernel or propagator can be expressed as the
product of a singular and a regular term, the latter of which
allows an expansion in a Taylor series, whose coefficients can
be obtained from the recursive solutions of a system
of ordinary differential equations. Their solutions simplify
considerably in the case of time-independent drift and diffusion
coefficients.
Indeed, with fully time-dependent drift and diffusion terms, each Taylor coefficient of the expansion depends on all the previous coefficients. 
Differently, with time-independent drift and diffusion terms, each coefficient only depends on the previous one.
This reduction in complexity, allows the procedure to be applied to cases of practical interest much more easily. 
Moreover, it can be applied manually in simple cases of theoretically interest and can be implemented in automatic symbolic computing environments in more complex cases of application interest. 

Subsequently we have put our newly developed machinery to work
on known examples of stochastic processes, highlighting specific
aspects of the implementation of our approach. 
Specifically, we have shown the application of our method to the simple case of Gaussian processes, to geometrical Brownian motions leading to the so-called log-normal distribution, and to a particular stochastic process used to describe chromatin remodeling phenomena.
In particular, we
see that the expansion cannot be guaranteed to work independently
of the discretization parameter $\alpha$, which we illustrated for the case of an exponential diffusion coefficient. We also showed
that in the cases in which the expansion works and leads to a
convergent expression for the time-dependent propagator - hence
also at long times - only a few of the expansion coefficients are
needed in order to already obtain a quantitatively reliable result.

Finally, we have taken advantage of our recursion system for the
determination of the expansion coefficients to allow us to find
novel drift and diffusion terms for which the procedure yields exact results for the propagator. This feature is of interest to be
further exploited to define novel stochastic processes of
interest in future applications.
Within the Fisk-Stratonovich interpretation, this approach allowed us to solve the stochastic differential equation without drift term. 
This result provides a unified picture for describing the geometric Brownian motion, the power law diffusion, and the exponential diffusion. 
These different heterogeneous diffusion schemes are well adapted to describe the anomalous diffusion observed in several physical and biological systems.
A further generalization concerns the Fisk-Stratonovich equation with drift.
Here we have proposed a technique that allows us to identify pairs of drift and diffusion terms for which it is easy to determine the exact expression of the propagator. This is useful for exploring new forms of stochastic equations and processes with possible various applications. 
From this point of view, an interesting future perspective concerns the application of this approach to problems with  drift or diffusion terms that are periodic in space. This can be useful to study diffusion processes in stratified or multilayered systems and to analyze the behaviors of reaction-diffusion equations with application to chemical processes and other physical instabilities and transformations.

\section{Appendix}

In this appendix we list, for completeness, the coefficients of
the development for the model discussed in Section \ref{secdna}, with the
results shown in Fig. \ref{dna}.

A first set of parameters is given by $k=2$, $\beta=1$, $\delta=1/10$, and $A=1$ (in arbitrary units) and the results have been obtained as
\begin{eqnarray}
D_0=1,
\end{eqnarray}
\begin{eqnarray}
 \nonumber 
D_1&=&2+\frac{1}{10}\,x-{\frac {67}{50}}\,{x}^{2}-{\frac {1099}{3000}}\,{x}^{3}-{
\frac {409}{10000}}\,{x}^{4}\\
&&-{\frac {91}{45000}}\,{x}^{5}+O(x^6),
\end{eqnarray}
\begin{eqnarray}
D_2&=&{\frac {13}{20}}+{\frac {1}{150}}\,x-{\frac {19969}{7500}}\,{x}^{2}-{
\frac {3251}{3750}}\,{x}^{3}+{\frac {1363801}{1750000}}\,{x}^{4} \nonumber\\
&&+{
\frac {50689969}{105000000}}\,{x}^{5}+O(x^6), 
\end{eqnarray}
\begin{eqnarray}
D_3&=&-{\frac {5033}{3750}}-{\frac {11641}{25000}}\,x+{\frac {1521223}{
2625000}}\,{x}^{2}+{\frac {4000063}{13125000}}\,{x}^{3}, \nonumber \\
&&+{\frac {
1149745319}{630000000}}\,{x}^{4}+{\frac {14432096843}{13500000000}}\,{
x}^{5}+O(x^6),
\end{eqnarray}
\begin{eqnarray}
D_4&=&-{\frac {3759293}{10500000}}-{\frac {16963}{750000}}\,x+{\frac {
1454450549}{315000000}}\,{x}^{2} \nonumber \\
&&+{\frac {11140823723}{4725000000}}\,{x
}^{3}-{\frac {57641437687}{63000000000}}\,{x}^{4} \nonumber \\
&&-{\frac {458475394523
}{506250000000}}\,{x}^{5}+O(x^6).
\end{eqnarray}

A second set of parameters is given by $k=2$, $\beta=1$, $\delta=1/10$, and $A=2$ (in arbitrary units) and we obtained
\begin{eqnarray}
D_0=1,
\end{eqnarray}
\begin{eqnarray}
 \nonumber 
D_1&=&2+\frac{1}{10}\,x-{\frac {17}{50}}\,{x}^{2}-{\frac {137}{1500}}\,{x}^{3}-{
\frac {307}{30000}}\,{x}^{4}\\
&&-{\frac {91}{180000}}\,{x}^{5}+O(x^6),
\end{eqnarray}
\begin{eqnarray}
 \nonumber 
D_2&=&\frac{3}{5}+{\frac {2}{75}}\,x-{\frac {10007}{15000}}\,{x}^{2}-{\frac {1621}{
7500}}\,{x}^{3}+{\frac {24527}{875000}}\,{x}^{4}\\
&&+{\frac {1524793}{
52500000}}\,{x}^{5}+O(x^6),
\end{eqnarray}
\begin{eqnarray}
 \nonumber 
D_3&=&-{\frac {2599}{1875}}-{\frac {1407}{3125}}\,x+{\frac {57521}{328125}}
\,{x}^{2}+{\frac {1992491}{26250000}}\,{x}^{3}\\
&&+{\frac {9657743}{
78750000}}\,{x}^{4}+{\frac {795208357}{11812500000}}\,{x}^{5}+O(x^6),
\end{eqnarray}
\begin{eqnarray}
 \nonumber 
D_4&=&-{\frac {165647}{656250}}-{\frac {7079}{93750}}\,x+{\frac {45878477}{
39375000}}\,{x}^{2}+{\frac {347956369}{590625000}}\,{x}^{3}\\
&&+{\frac {
741347219}{47250000000}}\,{x}^{4}-{\frac {45979054151}{885937500000}}
\,{x}^{5}+O(x^6).\,\,\,\,\,\,\,\,\,\,\,\,\,\,\,
\end{eqnarray}

\begin{acknowledgments}

S. G., F. C., and R. B. acknowledge
support funding of the French National Research Agency ANR through project `Dyprosome' (ANR-21-CE45-0032-02). 
\end{acknowledgments}


\begin{thebibliography}{100}

\bibitem{risken1989}
H. Risken, 
The Fokker-Planck Equation (Springer-Verlag, Berlin, 1989).

\bibitem{gardiner2009}
C. Gardiner, Stochastic Methods: A Handbook for the Natural and Social Sciences (Springer Verlag, Berlin 2009).

\bibitem{coffey2004}
W. T. Coffey, Yu. P. Kalmykov, and J. P. Waldron, The Langevin Equation (World Scientific, Singapore, 2004).

\bibitem{Silva2011}
A. T. Silva, E. K. Lenzi, L. R. Evangelista, M. K. Lenzi, H. V. Ribeiro,
and A. A. Tateishi, J. Math. Phys. \textbf{52}, 083301 (2011).

\bibitem{Cherstvy2013}
 A. G. Cherstvy, R. Metzler, Phys. Chem. Chem. Phys.  \textbf{15}, 20220 (2013).
 
\bibitem{Metzler2014}
R. Metzler, J.-H. Jeon, A. G. Cherstvy and E. Barkai, Phys. Chem. Chem. Phys.  \textbf{16}, 24128 (2014).

\bibitem{Cherstvy2017}
A. G. Cherstvy, D. Vinod, E. Aghion, A. V. Chechkin, and R. Metzler, New J. Phys. \textbf{19}, 063045 (2017).

\bibitem{Wang2020}
W. Wang, A. G. Cherstvy, X. Liu, and R. Metzler, Phys. Rev. E \textbf{102}, 012146 (2020).

\bibitem{Ritschel2021}
S. Ritschel, A. G. Cherstvy, and R. Metzler, J. Phys. Complex. \textbf{2}, 045003 (2021).

\bibitem{Vinod2022}
D. Vinod, A. G. Cherstvy, W. Wang, R. Metzler, and I. M. Sokolov, Phys. Rev. E \textbf{105}, L012106 (2022).

\bibitem{Vinod2022b}
D. Vinod, A. G. Cherstvy, R. Metzler, and I. M. Sokolov, Phys. Rev. E \textbf{106}, 034137 (2022).

\bibitem{ribeiro2023}
H. V. Ribeiro, A. A. Tateishi, E. K. Lenzi, R. L. Magin, and M. Perc, Communications Physics \textbf{6}, 244 (2023). 
 
\bibitem{leibovich2019}
N. Leibovich and E. Barkai, 
Phys. Rev. E \textbf{99}, 042138 (2019).

\bibitem{aghion2019}
E. Aghion, D. A. Kessler, and E. Barkai, Phys. Rev. Lett. \textbf{122}, 010601 (2019).

\bibitem{aghion2020} 
E. Aghion, D. A. Kessler, and E. Barkai, Chaos, Solitons and Fractals \textbf{138}, 109890 (2020).
 
\bibitem{giordano2023}
S. Giordano, F. Cleri, and R. Blossey, 
Phys. Rev. E \textbf{107}, 044111 (2023).
 
\bibitem{drozdov1993a} A. N. Drozdov, 
Physica A \textbf{196}, 258-282 (1993).

\bibitem{drozdov1993} A. N. Drozdov, 
Physica A \textbf{196}, 283-312 (1993).

\bibitem{drozdov1995} A. N. Drozdov, 
Phys. Rev. Lett. \textbf{75}, 4342-4345 (1995).
 
\bibitem{weiss1995} G. H. Weiss and M. Gitterman,
Phys. Rev. E \textbf{51}, 122-125 (1995).

\bibitem{drozdov1996} A. N. Drozdov and M. Morillo,
Phys. Rev. Lett. \textbf{77}, 3280-3283 (1996)  .

\bibitem{donoso1999}
M. Donoso, J. J. Salgado and M. Soler, J. Phys. A: Math. Gen. \textbf{32},  3681–3695 (1999).

\bibitem{nistor2010}
R. Constantinescu, N. Costanzino, A. L. Mazzucato, and V. Nistor, J. Math. Phys. \textbf{51}, 103502 (2010).

\bibitem{nistor2011}
W. Cheng, N. Costanzino, J. Liechty, A. Mazzucato, and V. Nistor, SIAM J. Financial Math. \textbf{2}, 901-934 (2011).


\bibitem{bilal2020} A. Bilal, 
J. Math. Phys. \textbf{61}, 061517 (2020).

\bibitem{kumar2008}
K. V. Kumar, S. Ramaswamy, and M. Rao, Active elastic dimers: Self-propulsion and current reversal on a featureless track, Phys. Rev. E 77, 020102(R) (2008).

\bibitem{baule2008} A. Baule, K.V. Kumar, S. Ramaswamy, 
J. Stat. Mech. \textbf{2008} P11008 (2008).

\bibitem{blossey2019} R. Blossey, H. Schiessel, 
J. Phys. A: Math. Theor. {\bf 52}, 085601 (2019).

\bibitem{bauermann2019} J. Bauermann, B. Lindner,
BioSystems {\bf 178}, 25-31 (2019).

\bibitem{zhu2021}
Z. Zhu, G. Ren, X. Zhang, J. Ma, Chaos, Solitons and Fractals \textbf{151},  111203 (2021).

\bibitem{liu2004}
Q. Liu, Y. Jia, Phys. Rev E \textbf{70}, 041907 (2004).

\bibitem{frigola2012}
D. Frigola, L. Casanellas, J. M. Sancho, M. Iba\~{n}es, PLoS ONE \textbf{7}, e31407 (2012).

\bibitem{manca2016}
F. Manca, P.-M. Déjardin, and S. Giordano, Ann. Phys. (Berlin) \textbf{528}, 381 (2016).

\bibitem{giordano2019}
S. Giordano, Eur. Phys. J. B \textbf{92}, 174 (2019).

\bibitem{giordano2021}
S. Giordano, Phys. Rev. E \textbf{103}, 052116 (2021).

\bibitem{landi2014}
G. T. Landi  and M. J. de Oliveira, Phys. Rev. E \textbf{89}, 022105 (2014).

\bibitem{palla2020} 
P. L. Palla, G. Patera, F. Cleri, and S. Giordano, Physica Scripta \textbf{95}, 075703 (2020).

\bibitem{fuchs2022} A. Fuchs, C. Herbert, J. Rolland, M. W\"achter, F. Bouchet, J. Peinke,
Phys. Rev. Lett. \textbf{129}, 034502 (2022).

\bibitem{birnir2013} 
B. Birnir, The Kolmogorov-Obukhov theory of turbulence (Springer, New York, 2013).

\bibitem{coffey1985}
W. Coffey, Adv. Chem. Phys. \textbf{63}, 69 (1985).

\bibitem{oksendal2003} B. {\O}ksendal,
Stochastic differential equations (Springer Berlin, 2003).

\bibitem{ito1950}
 K. It\^{o}, Nagoya Math. J. \textbf{1}, 35 (1950).
 
\bibitem{fisk1963}
D. L. Fisk,  Quasi-martingales and stochastic integrals (Kent: Research monograph, Kent State University, 1963).

\bibitem{stratonovich1966}
R. L. Stratonovich, SIAM J. Control Optim. \textbf{4}, 362 (1966).
 
\bibitem{haenggi1982}
P. H\"{a}nggi, H. Thomas, Phys. Rep. \textbf{88}, 207 (1982).
 
\bibitem{klimontovich1995}
Yu. L. Klimontovich, Statistical Theory of Open Systems (Kluver Academic, Dordrecht, 1995).

\bibitem{sokolov2010} 
I. M. Sokolov, Chem. Phys. \textbf{375}, 359 (2010).

\bibitem{denisov2009} 
S. I. Denisov, W. Horsthemke, P. H\"anggi,
Eur. Phys. J. B {\bf 68}, 567-575 (2009).

\bibitem{denisov2003}
S. I. Denisov, A.N. Vitrenko, W. Horsthemke, 
Phys. Rev. E \textbf{68}, 046132 (2003).
 
\bibitem{denisov2014}
V. M\'{e}ndez, S.I. Denisov, D. Campos, W. Horsthemke, 
Phys. Rev. E \textbf{90}, 012116 (2014).

\bibitem{uhlenbeck1930}
G. E. Uhlenbeck and L. S. Ornstein, 
Phys. Rev. \textbf{36}, 823 (1930).

\bibitem{wang1945}
M. C. Wang and G. E. Uhlenbeck, Rev. Mod. Phys. \textbf{17}, 323 (1945).

\bibitem{mantegna2000} 
R. N. Mantegna, H.E. Stanley,
An Introduction to Econophysics (Cambridge University Press, 2000).

\bibitem{bouchaud2009} 
J.-P. Bouchaud, M. Potters, Theory of Financial Risk and Derivative Pricing (Cambridge University Press, 2009).

\bibitem{hull2021} 
J. C. Hull, Options, Futures, and Other Derivatives (Pearson Education Ltd., 2021).


\bibitem{Stojkoski2020}
V. Stojkoski, T. Sandev, L. Basnarkov, L. Kocarev, and R. Metzler, Entropy \textbf{22}, 1432 (2020).


\bibitem{saaty1981}
T. L. Saaty, Modern nonlinear equations (Dover Publications, New York, 1981).

\bibitem{hapca2007}
S. Hapca, J. Crawford, R. Rae, M. Wilson, and I. Young, Biological Control \textbf{41}, 223–229 (2007).

\bibitem{hapca2007b}
S. Hapca, J. W. Crawford, K. MacMillan, M. J. Wilson and I. M. Young, J. Theor. Biol. \textbf{248}, 212 (2007).

\bibitem{hapca2009}
S. Hapca, J. W. Crawford and I. M. Young, J. R. Soc., Interface \textbf{6}, 111 (2009).

\bibitem{buste2010}
S. B. Yuste, E. Abad and K. Lindenberg, Phys. Rev. E \textbf{82}, 061123 (2010).

\bibitem{crick1970}
F. Crick, Nature (London) \textbf{225}, 420 (1970).

\bibitem{kowall1976}
J. Kowall, D. Peak and J. W. Corbett, Phys. Rev. B: Condens. Matter Mater. Phys., 1976, \textbf{13}, 477.

\bibitem{kesarev2009}
A. G. Kesarev and V. V. Kondrat’ev, Phys. Met. Metallogr. \textbf{108}, 30 (2009).

\bibitem{kesarev2015}
A. G. Kesarev, V. V. Kondrat’ev, and I. L. Lomaev, Phys. Met. Metallogr. \textbf{116}, 225–234 (2015).

\bibitem{fa2005}
K. S. Fa, Phys. Rev. E  \textbf{72}, 020101(R) (2005).
 
\bibitem{kimura1964}
M. Kimura, Journal of Applied Probability \textbf{1}, 177-232 (1964).

\bibitem{montroll1971}
N. S. Goel, S. C. Maitra, and E. W. Montroll, Rev. Mod. Phys. \textbf{43}, 231-276 (1971).

\bibitem{birch1999}
C. P. D. Birch, Annals of Botany \textbf{83}, 713–723 (1999).

\bibitem{sakanoue2007}
S. Sakanoue, Ecological Modelling \textbf{205}, 159-168 (2007). 

\bibitem{fa2018}
K. S. Fa, Langevin and Fokker-Planck equations and their generalizations (World Scientific, Singapore, 2018).
 






















\end{thebibliography}
\end{document}